\documentclass[%
 reprint,
 amsmath,amssymb,
 aps,
floatfix,
]{revtex4-2}

\usepackage{graphicx}% Include figure files
\usepackage{dcolumn}% Align table columns on the decimal point
\usepackage{bm}% bold math
\usepackage{subfigure}
\usepackage{wrapfig}
\usepackage{multirow}
\usepackage{hyperref}% add hypertext capabilities
\usepackage{amsmath}
\usepackage{mleftright}
\usepackage{lipsum}
\usepackage{amstext}
\usepackage{xcolor}
\usepackage{mwe}
\usepackage[export]{adjustbox}

\newcommand\scalemath[2]{\scalebox{#1}{\mbox{\ensuremath{\displaystyle #2}}}}

\begin{document}

\preprint{APS/123-QED}

\title{Enhancing Quantum Otto Engine Performance in \\ Generalized External Potential on Bose-Einstein Condensation Regime}% Force line breaks with \\
 
\author{Zahara Zettira$^1$}
\author{Ade Fahriza$^1$}
\author{Zulfi Abdullah$^1$}
\author{Trengginas E P Sutantyo$^1$}
\email{trengginasekaputra@sci.unand.ac.id}
\affiliation{%
$^1$Theoretical Physics Laboratory, Department of Physics, Faculty of Mathematics and Natural Science, Andalas University, Indonesia}%
\date{\today}% It is always \today, today,

\begin{abstract}
We examine a quantum Otto engine using both Bose-Einstein Condensation (BEC) and normal Bose gas as working medium trapped in generalized external potential. We treated the engine quasi-statically and endoreversibly. Since the expansion and compression in both quasi-static and endoreversible take place isentropic, the expression of efficiency is similar. However, the power output in the quasi-static cycle is zero due to infinite and long stroke time. In contrast, with an endoreversible cycle, thermalization with two reservoirs takes place at a finite time. We use Fourier’s law in conduction to formulate the relation between temperature of medium and reservoir, making work depend on heating and cooling stroke time. Moreover, we maximized the power with respect to compression ratio $\kappa$ to obtain efficiency at maximum power (EMP). We found that EMP is significantly higher when using BEC as a working medium, meanwhile EMP with normal Bose gas is just Curzon-Ahlborn efficiency. We also investigate the effect of thermal contact time $\tau$ with hot $(\tau_{h})$ and cold $(\tau_{l})$ reservoir on EMP. We found that when complete thermalization, $\tau_{h}=\tau_{l}$, stroke time occurs, there are no significant differences. Nevertheless, while incomplete thermalization arise, by adjusting various cooling and heating stroke time, provides a significant result on EMP, which is much higher at $\tau_{h}<\tau_{l}$ stroke time whilst lower at $\tau_{h}>\tau_{l}$ stroke time. We conclude this incomplete thermalization leads to the condition where residual coherence emerges which enhances the EMP of the engine.
\end{abstract}

\maketitle

%\tableofcontents

\section{Introduction}

Nowadays, quantum science significantly influences the development of classical physics theory, unexceptionable thermodynamics, which has recently been known as quantum thermodynamics \cite{Deffner.2019} or nano thermodynamics \cite{Chamberlin.2015}. Since the discovery of quantum science, nano thermodynamics has emerged rapidly \cite{Dong.2023, Pena.2023, Galteland.2021, Strom.2020, Miguel.2020, YYin.2018, Strom.2017, Kjelstrup.2014} and being fundamental for future advanced technologies \cite{KJinuk.2022}; quantum information, high precision sensors, and quantum heat engine (QHE). Started by the pioneer of QHE \cite{HEDScovil.1959} until recent development \cite{KJinuk.2022, KMoochan.2022}; this revolution is a bridge to the breakthrough implementation of nano heat engines. Over the last two decades, research on quantum heat
engines has opened up the horizons of their benefits.  \cite{MyersAbah.2022}. The motivation is to realize the QHE as close as possible to the practical world as a worthwhile device.

Quantum Heat Engine (QHE) is a device that utilizes quantum matter or particles as its working medium in order to convert heat into work \cite{HTQuan.2007}. In fact, the QHE also operates using a cycle that resembles classical thermodynamics, such as the Otto \cite{Papadatos.2021, LLi.2021, SSingh1.2020, OAbah.2019, FAltintas.2019}, Lenoir \cite{Fahriza.2022, Fahriza2.2022, FAbdillah.2021, YDSaputra.2021, MHAhmadi.2019, RWang.2012}, Diesel \cite{SSingh3.2020, DPSetyo.2018}, and Carnot \cite{CMBender.2000, CMBender.2002, FMoukalled.1995, FAltintas.2019, IHBelfaqih.2015, TEPSutantyo.2020, TEPSutantyo.2015} cycles. Nevertheless, the physical quantities in the quantum thermodynamic cycle are different from the classical thermodynamic cycle; this is the underlying reason that the efficiency of a quantum heat engine is ably outperformed the most efficient classical heat engine, the Carnot Engine \cite{Scully.2003}. In a certain state, the work produced by QHE exceeds the maximum value of a heat engine which operates classically \cite{Deffner.2018, TDKieu.2006, TDKieau.2004}.

A classical heat engine that operates with a reversible cycle produces the highest efficiency, e.g., classical Carnot engine. However, the Carnot engine operates via a quasistatic process that lasts long and produces zero power output, making it impossible to realize. Curzon and Ahlborn \cite{CA.1975} applied an endoreversible approach, in which the process is accelerated, and the power output produced by the engine is finite. Although the efficiency $\eta_{CA}$ is lower than the efficiency of a quasistatic Carnot engine $\eta_C$, an endoreversible Carnot engine is more realistic and possible to realize.

Recently, many heat engine models proposed to investigate the endoreversible cycle because it is more realistic to implement \cite{Abah.2012, Deffner.2018, Kosloff.2017, Myers.2022, MyersDeffner.2020, Ross.2014, Smith.2020, Zettira.2023}, even those endoreversible heat engines have an efficiency that exceeds Curzon-Ahlborn limit \cite{CA.1975}. The endoreversible heat engine model uses a variety of mediums such as classical gas \cite{Erbay.1997, Leff.1987}, single ion \cite{Abah.2012,Deffner.2018,Ross.2014}, or even quantum matter in the form of fermions \cite{Myers1.2022} and bosons \cite{Fialko.2012, Gluza.2021, Li.2022, MyersDeffner.2020}.

Interestingly, boson as a working medium produces greater efficiency than fermion achieves \cite{MyersDeffner.2020}, due to the advantages of symmetry possessed by boson that fermion does not. This is what underlies this research to examine the role of boson as a working medium, especially in the Bose-Einstein Condensation (BEC) regime, which is even claimed to be able to provide better performance than ordinary boson gas \cite{Myers.2022}. The use of BEC as a heat engine working medium is being intensively investigated \cite{Fialko.2012, Gluza.2021, Li.2022, Li.2018, Myers.2022}. When the bosonic atoms gas is cooled below a certain critical temperature $T_c$, such that most of the atoms are condensed in the lowest quantum state, experience a phase transition into BEC regime \cite{Einstein.1925, Bose.1924}. BEC is a quantum phenomenon that can be realized on several vapors, i.e., rubidium \cite{Anderson.2008}, sodium \cite{Davis.1995}, and lithium \cite{Bradley.1997, Bradley.1995}. Moreover, recently BEC can be observed in a 3D potential box  \cite{Gaunt.2013}. 

In this study, we examine the quantum Otto engine using the advantage of the BEC's thermodynamic properties as a working medium under the influence of a 3D generalized external trapping potential. We modify that typical potential \cite{Bagnato.1987} in order to investigate performance in other circumstances, especially considering the possibilities of the cycles that can exploit external trapping potential as a bridge to extending the analysis to the nonequilibrium regime \cite{Myers.2022}. Besides, with that potential, the engine aims to obtain better performance, considering for $n\rightarrow \infty$, i.e., box potential, provides lower heat capacity $(c_{V})$ than for $n=2$, i.e., harmonic potential \cite{Bagnato.1987}. This will affect the efficiency of the Otto engine $\eta(c_V)$, with box potential achieving higher efficiency than harmonic potential. In addition, we compare the performance of quantum Otto engine from work and efficiency both in condensed and non-condensed phase. Apart from that, we investigate the engine performance from efficiency at maximum power (EMP) by maximizing the power output to the compression ratio parameter. Lastly, we also explore the engine performance by varying thermalization time during the cycles, which at a specific value boosts the EMP significantly.

\section{Thermodynamic Properties of BEC in Generalized External Potential}\label{sec2} 

\begin{figure}[!b]
\centering
\includegraphics[width=.22\textwidth]{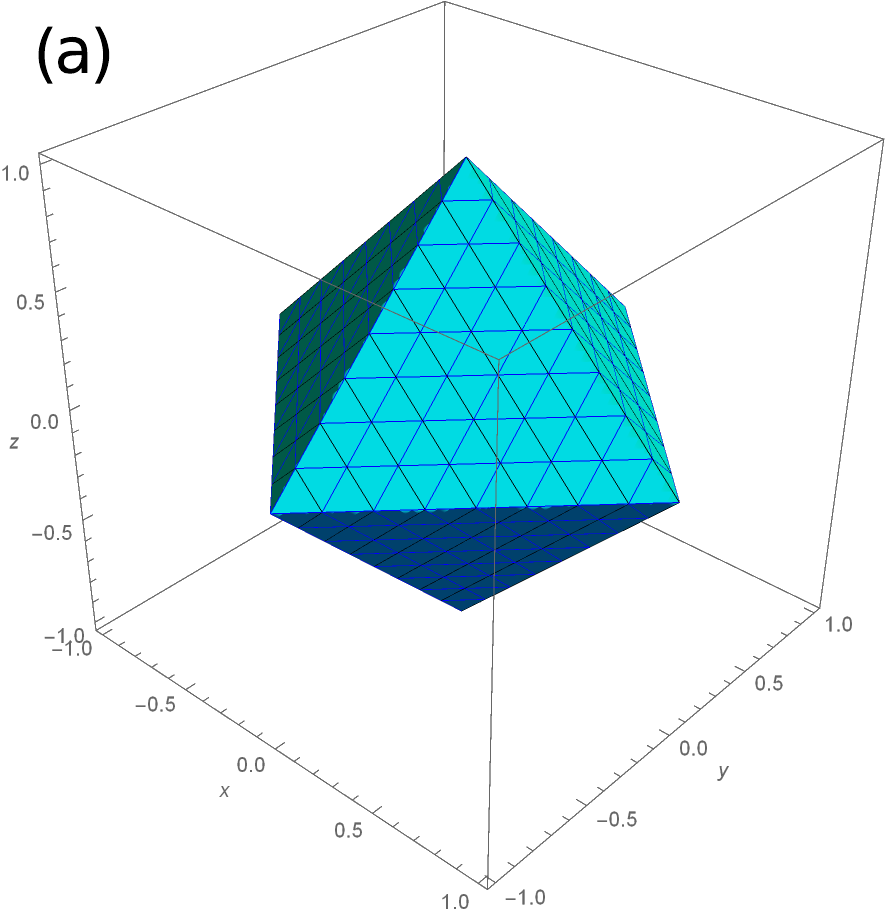}\hfill
\includegraphics[width=.22\textwidth]{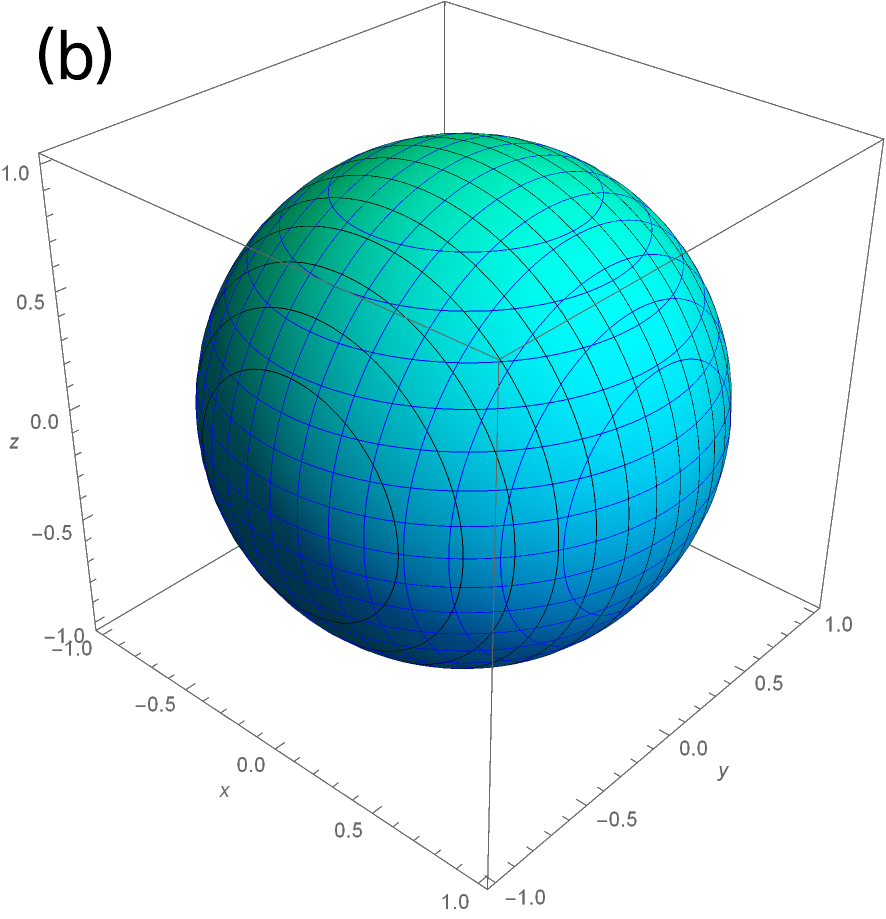}\hfill
\vspace{0.4cm}
\includegraphics[width=.22\textwidth]{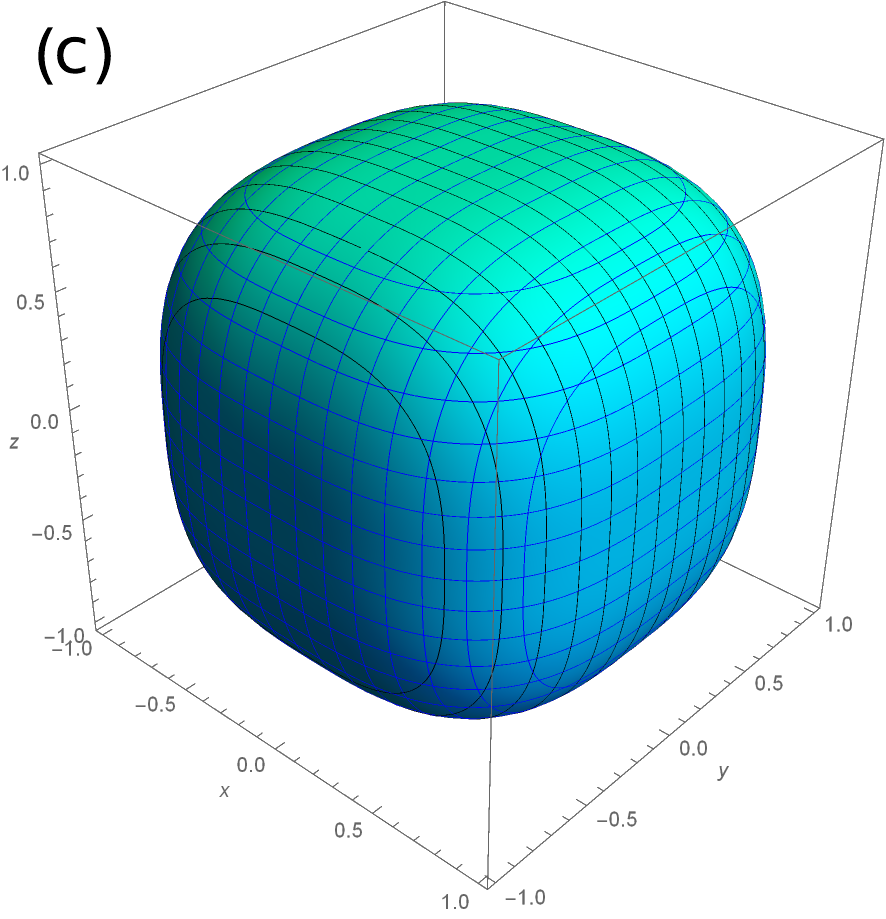}\hfill
\includegraphics[width=.22\textwidth]{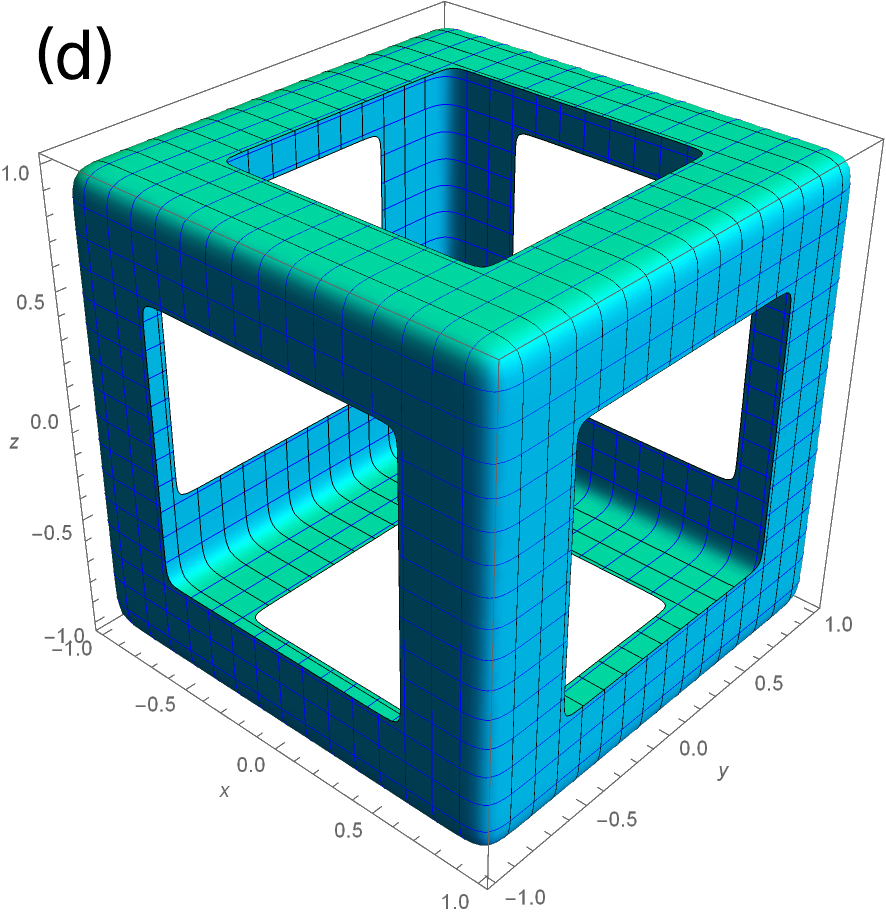}
\caption{\footnotesize\label{fig1} Illustration of external potential with (a) $n = 1$, (b) $n = 2$, (c) $n = 3$, \text{ and } (d) $n \rightarrow \infty$ respectively.}	
\end{figure}

We consider $N$ number of particles in an ideal Bose gas to be equally distributed in a certain energy level. This energy level is determined by the Hamiltonian worked in the aforementioned particles. %As we know, Hamiltonian consists of kinetic energy $K$ and potential energy $V$.% 
In this study, we define ideal Bose gas in generalized external trapping potential as formulated below
\begin{align}
    V(\boldsymbol{r})= \varepsilon_{0} \left( \left|\frac{x}{a} \right|^{p} + \left|\frac{y}{a}\right|^{q} + \left| \frac{z}{a}\right|^{l} \right) \label{eq1}
\end{align}
where $x$, $y$, and $z$ are coordinate of the particles' positions in space, $\varepsilon_{0}$ is a constant in energy dimension, and $a$ is radius of the potential. Here, we set $p = q = l = n$, so for certain $n$ numbers, the potential has shape as Figure \ref{fig1}. 

The density of state for distributed particles in potential from Equation \ref{eq1} is given by \cite{Bagnato.1987}
\begin{align}
    \rho(\varepsilon) =  \scalemath{0.94}{\left[ \frac{2\pi(2m)^{\frac{3}{2}}}{h^{3}} \right]} \frac{a^{3}}{\varepsilon_{0}^{\frac{3}{n}}} \varepsilon^{\lambda} F(n,n,n) \label{eq2}
\end{align}
where $\lambda  
%= \frac{1}{n} + \frac{1}{n} + \frac{1}{n} + \frac{1}{2} 
= \frac{3}{n} + \frac{1}{2}$ and $F(n,n,n)$ is defined as below 
\begin{align}
    \scalemath{1}{F(n,n,n) =}
    &\scalemath{0.83}
    {\biggl[ \int_{-1}^{1} (1 - X^{n})^{\frac{1}{2}+\frac{2}{n}} \,dX \biggr]} \times \nonumber \\ 
    &\scalemath{0.83}
    {\biggl[ \int_{-1}^{1} (1 - X^{n})^{\frac{1}{2}+\frac{1}{n}} \,dX \biggr]} \times \nonumber \\
    &\scalemath{0.83}
    {\biggl[ \int_{-1}^{1} (1 - X^{n})^{\frac{1}{2}} \,dX \biggr]}.
    \label{eq3}
\end{align}

Grand potential for the boson particle system is given by \cite{Pathria.2011}
\begin{align}
    \Omega = k_{B}T \sum_{i} \ln\left(1-ze^{-\beta\varepsilon_{i}}\right) \label{eq4}
\end{align}
which $z = e^{\frac{\mu}{k_{B}T}}$ is the fugacity. Since in BEC cases, the big fraction of bosons are condensed into ground state energy, we can consider $k_{B}T \gg \varepsilon_{i+1}-\varepsilon_{i}$. Therefore, the system can be defined as a continuous state plus a discrete ground state. By substituting Equation \ref{eq2} to Equation \ref{eq4}, we obtain
\begin{align}
    \Omega &= Ak_{B}Ta^{3} \int^{\infty}_{0} \varepsilon^{\frac{3}{n}+\frac{1}{2}} \ln\left(1-ze^{-\varepsilon\beta}\right)d\varepsilon \nonumber \\
    &= -A'a^{3}(k_{B}T)^{\frac{3}{n}+\frac{5}{2}} g_{\frac{3}{n}+\frac{5}{2}} (z) \label{eq5}
\end{align}
with $A = \left[ \frac{2\pi(2m)^{\frac{3}{2}}}{h^{3}} \right] \frac{F(n,n,n)}{\varepsilon_{0}^{\frac{3}{n}}}$ and $A' = \frac{1}{\frac{3}{n}+\frac{3}{2}} \Gamma \left( \frac{3}{n} + \frac{5}{2} \right)A$ which are constants and $b=\frac{3}{n}+\frac{3}{2}$ is used to shorten the notations. Furthermore, $g_{\frac{3}{n}+\frac{5}{2}}(z)$ is the related Bose function which is defined in the following equation \cite{Pathria.2011}
\begin{align}
    g_{v}(z) = \frac{1}{\Gamma(v)} \int^{\infty}_{0} dx \frac{1}{z^{-1}e^{x}-1} \label{eq6}
\end{align}
Bose function can also be defined as a series that is similar to the Zeta function series,
\begin{align}
    g_{v}(z) = \sum^{\infty}_{n=1} \frac{z^{n}}{n^{v}} \label{eq7}
\end{align}
and can also be expressed in the following recursion relation,
\begin{align}
    g_{v-1}(z) = \frac{\partial}{\partial \ln(z)} g_{v} (z) \label{eq8}
\end{align}

Based on $\Omega = U - TS - \mu N$ relation \cite{Pathria.2011}, we can derived $N$ that is the number of excited atom at temperature $T$ 
\begin{align}
    N = -\biggl( \frac{\partial\Omega}{\partial\mu} \biggr)_{T,a} = A'a^{3}(k_{B}T)^{b} g_{b} (z) \label{eq9}
\end{align}
BEC occurs when $\mu$ increases monotonously until the value approaches the energy at ground state ($\varepsilon_{0}$). If $\varepsilon_{0} = 0$ \cite{Pita.2016}, then BEC occurs at $\mu = 0$ or $z = 1$ so that we can derive the critical temperature, the temperature at the phase transition, as
\begin{align}
    T_{c} = \left[ \frac{N}{A'a^{3}}\frac{1}{\zeta(b)} \right]^{\frac{1}{b}}\frac{1}{k_{B}}
    \label{eq10}
\end{align}
The entropy $S$ and the internal energy $U$ of the BEC can be derived respectively, as follow
\begin{widetext}
\begin{align}
    S = -\biggl( \frac{\partial\Omega}{\partial T} \biggr)_{T,a,\mu} =
    \begin{cases}
    A'a^{3}k_{B}(k_{B}T)^{b} \left[ (b+1) g_{b+1} (1) \right] & T \le T_{c} \\
    A'a^{3}k_{B}(k_{B}T)^{b} \left[ (b+1) g_{b+1} (z) - \ln zg_{b} (z) \right] & T \ge T_{c}
    \end{cases}
    \label{eq11}
\end{align}
and
\begin{align}
    U = \Omega + TS + \mu N =
    \begin{cases}
    (b) A'a^{3}(k_{B}T)^{b+1} g_{b+1} (1) & T \le T_{c} \\
    (b) A'a^{3}(k_{B}T)^{b+1} g_{b+1} (z) & T \ge T_{c}
    \end{cases}
    \label{eq12}
\end{align}
\end{widetext}

\section{Endoreversible Otto Cycle}\label{sec3} 

The main principle of endoreversible thermodynamics is local equilibrium. The transformation of working medium during isentropic stroke takes place slowly so that the working medium is always able to reach equilibrium. However, because the cycle time is finite, the working medium does not have time to reach equilibrium with the reservoir. So, from the reservoir point of view, the process is irreversible, but from the working medium point of view itself, the process is reversible \cite{Hoffman.1997}.

\begin{figure}[!htp]
\includegraphics[width=0.45\textwidth]{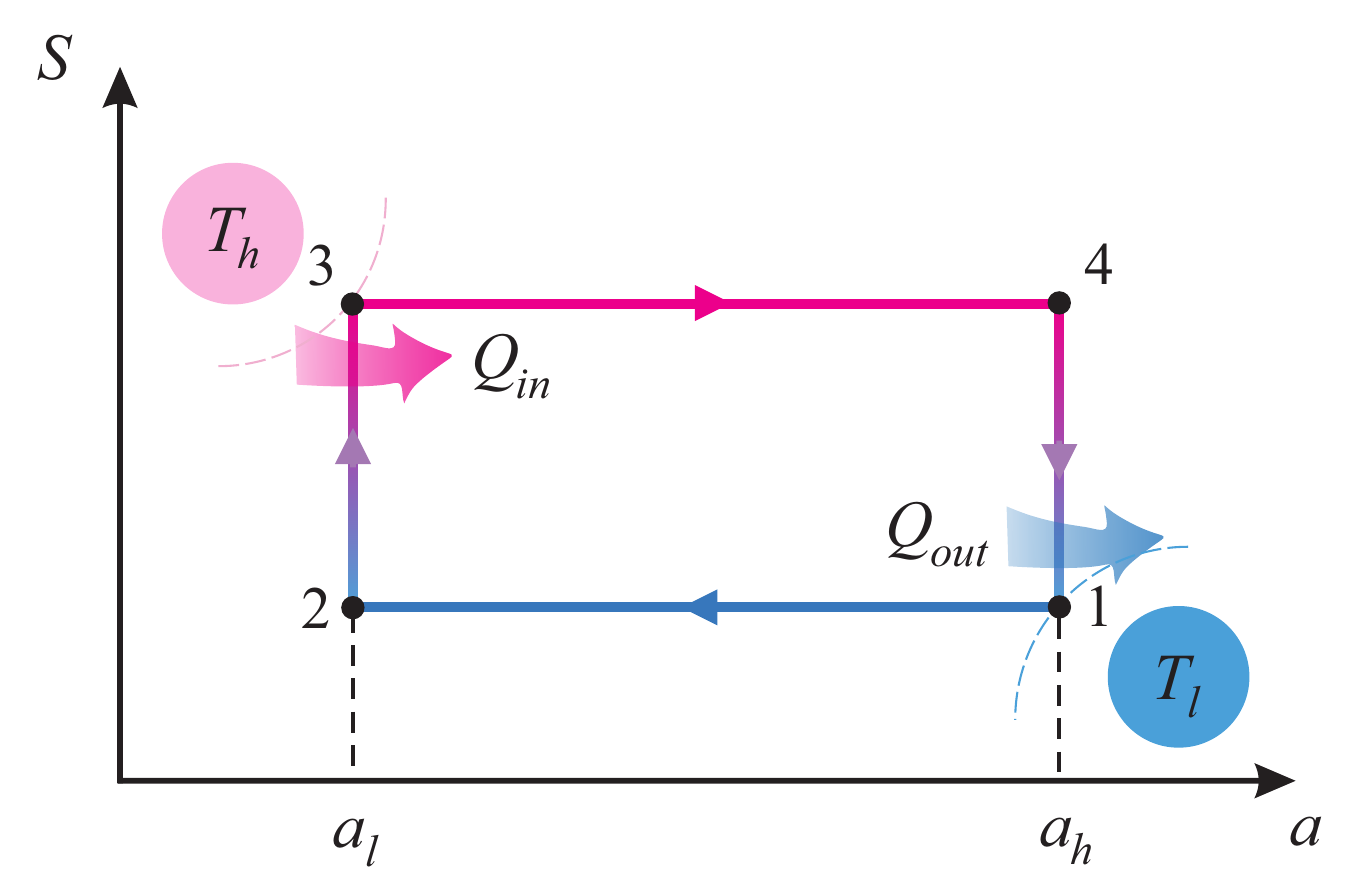}
\caption{\footnotesize\label{fig2} Otto cycle diagram for Entropy $S$ vs. external field $a$ which heat exchange occurs only between 4-1 and 2-3 strokes due to isolation of isentropic conditions through 1-2 and 3-4 strokes.}
\centering
\end{figure}

An ideal Otto cycle consists of four strokes: isentropic compression (stroke 1-2), isochoric heating (stroke 2-3), isentropic expansion (stroke 3-4), and isochoric cooling (stroke 4-1) \cite{Cengel.2008}, as shown in Figure \ref{fig2}. By varying $a$ as a parameter shows potential radius, it means we variate the external potential for the system which also means there is a change in order of the energy at said system until the system is excited. On the other hand, imbuing thermal energy into the system can also cause excitation in the system. The energy transfer of the external field and the thermal energy are arranged in such a way that the Otto cycle is relevant to the classical Otto cycle.

During isochoric heating stroke, external field remains constant at $a_{l}$ so as the volume and the system make contact with the hot reservoir, the heat flows into the system. Based on the first law of Thermodynamic, the amount of heat transferred is
\begin{align}
    \Delta Q_{in} = U \left(T_{3}, a_{l}\right) - U \left(T_{2}, a_{l}\right) 
    \label{eq13}
\end{align}
Heat transfer during heating stroke, the system obeys the Fourier heat conduction law as given
\begin{align}
    \frac{dT}{dt} = -\alpha_{h}\left( T-T_{h}\right)
    \label{eq14}
\end{align}
where $\alpha_{h}$ is a constant that depends on thermal conductivity and thermal capacity of the medium during heating stroke. The boundary conditions at the beginning and the end of isochoric stroke are written as
\begin{align}
    T(0) = T_{2} \text{ and } T(\tau_{h}) = T_{3}, \text{ where } T_{2} < T_{3} \leq T_{h}
    \label{eq15}
\end{align}
with solution
\begin{align}
    T_{3} - T_{h} = (T_{2} - T_{h}) e^{-\alpha_{h}\tau_{h}}
    \label{eq16}
\end{align}
when a time limit of heating stroke goes to infinity, $\tau_{h} \rightarrow \infty$, $T_{3} = T_{h}$ and it becomes the quasi-static condition. 

During isentropic expansion stroke, external field is varied from $a_{l}$ to $a_{h}$ which also means the volume of the system is changed from $V_{l}$ to $V_{h}$. However, the system is unattached from the reservoir so that there is no heat exchange within the system $(\Delta Q = 0)$. From the First Law of Thermodynamics, we get
\begin{align}
    W_{exp} =  U\left(T_{4}, a_{h}\right) - U\left(T_{3}, a_{l}\right) 
    \label{eq17}
\end{align}

After the volume reach $V_{h}$, the system is back interconnected to the cold reservoir. During this stroke, the external field remains constant at $a_{h}$. Similar to the isochoric heating stroke, the ejected heat is the internal energy difference from stroke 3 to 2,
\begin{align}
    \Delta Q_{out} = U \left(T_{1}, a_{h}\right) - U \left(T_{4}, a_{h}\right) 
    \label{eq18}
\end{align}
Heat transfer process during cooling stroke which also obey the Fourier heat conduction law as below
\begin{align}
    \frac{dT}{dt} = -\alpha_{l}\left( T-T_{l}\right)
    \label{eq19}
\end{align}
where $a_{l}$ is a constant of thermal conductivity and thermal capacity of the medium during cooling stroke. This isochoric stroke has boundary conditions as
\begin{align}
    T(0) = T_{4} \text{ and } T(\tau_{l}) = T_{1}, \text{ where } T_{4} > T_{1} \geq T_{l}
    \label{eq20}
\end{align}
with solution
\begin{align}
    T_{1} - T_{l} = (T_{4} - T_{l}) e^{-\alpha_{l}\tau_{l}}
    \label{eq21}
\end{align}
when quasi-static condition is achieved while cooling stroke time reaches the limit to infinity, $\tau_{l} \rightarrow \infty$, in other words, $T_{1} = T_{l}$.

Furthermore, the system is re-disconnected from the reservoir and the field is varied $a_{h}$ to $a_{l}$ to fulfill isentropic conditions. As there is no heat transfer within the system, then the work during compressing stroke is given by
\begin{align}
    W_{comp} =  U\left(T_{2}, a_{l}\right) - U\left(T_{1}, a_{h}\right) 
    \label{eq22}
\end{align}
Unlike the quasi-static Otto cycle, in which the four strokes are reversible, in the endoreversible Otto cycle, 2 strokes are irreversible and 2 strokes are reversible. Irreversibility occurs when the working medium comes into contact with the reservoir during isochoric stroke, which causes leaking during the heat transfer \cite{JWang.2012}. On the other hand, during isentropic strokes, there is no heat leaking occurs due to the system being isolated to external energy \cite{JWang.2012}. This idea has been demonstrated in recent studies. \cite{Deffner.2018, Myers.2022, Smith.2020, Wang.2009} 

Next, we determine the thermal efficiency by comparing the total work done in a cycle with the heat in
\begin{align}
    \eta = -\frac{W_{exp}+W_{comp}}{Q_{in}} 
    \label{eq23}
\end{align}
whilst the power output is determined by the total work done and the time for one cycle
\begin{align}
    P = -\frac{W_{exp}+W_{comp}}{\gamma{\tau_{h}+\tau_{l}}} 
    \label{eq24}
\end{align}
with $\gamma$ is a multiplier constant that refers to the total time for one cycle, including the time throughout isentropic strokes.

To eliminate $T_{1}, T_{2}, T_{3}, \text{and} T_{4}$ into controllable parameters ($T_{h}, T_{l}$), both Equation \ref{eq17} and \ref{eq22} are not enough so that another two equations are needed. During isentropic strokes, the entropy and fugacity are constant \cite{Myers.2022, Deffner.2018} so based on Equation \ref{eq11}, within isentropic strokes, we got the relation of temperature as
\begin{align}
    T_{2} = \kappa^{-\frac{1}{b}} T_{1} \text{ and } T_{4} = \kappa^{\frac{1}{b}} T_{3}
    \label{eq25}
\end{align}
which $\kappa = \left( \frac{a_{l}}{a_{h}} \right)^3$ is the ratio of volume compression.

\section{Result and Discussion} \label{sec4}
\subsection{Quasi-static Performance} \label{sec4.1}

By using Equation \ref{eq11}, we obtain the critical temperature for $n = 1,2,3 \text{ and } n \rightarrow \infty$, as shown in table 1 that $T_{c}$ decreases as $n$ increases. $T_{c}$ for $n = 2$ is the same as Myers et al. obtained in their research \cite{Myers.2022} and other related data have also been adjusted according to reference \cite{Aveline2020}. This decrease in $T_{c}$ is related to the volume of each potential as shown in Figure \ref{fig1}, where volume is proportional to $n$. Thus at the same number of particles $N$, potential with small $V$, the density of atoms is greater than the potential with large $V$. This is consistent with the results obtained by \cite{Reppy.2000} that an increase in density of atoms causes an increase in $T_{c}$ as well, so that at  $n \rightarrow \infty$ the gas becomes less dense so $T_{c}$ is also small.

\begin{table}[htp]
\begin{ruledtabular}
\caption{\footnotesize Thermodynamics property of BEC using potential $n = 1, 2, 3 \text{ and } n \rightarrow \infty$ with 60.000 bosons (Rubidium-87) and $a = 1 \text{ mm}$} \label{tab1}%
\begin{tabular}{cccccc}
$n$    & $F(n,n,n)$   & $b$  & $T_{c}$ (nK)  & $C_{v}(T^{-}_{c})/Nk_{B}$ &  $C_{p}(T^{-}_{c})/Nk_{B}$ \\
\hline
1         & 13.78    & 9/2   & 280.39   & 24.06    & 29.41 \\
2         & 2.46     & 3     & 41.28    & 10.82    & 14.40 \\
3         & 7.83     & 5/2   & 6.05     & 7.34     & 10.28 \\
$\infty$  & 8        & 3/2   & 0.00722  & 1.92     & 3.20 \\
\end{tabular}
\end{ruledtabular}
\end{table}

\begin{figure}[!t]
\includegraphics[width=0.48\textwidth]{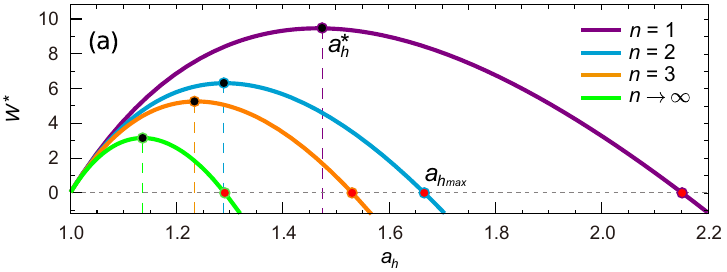}
\includegraphics[width=0.48\textwidth]{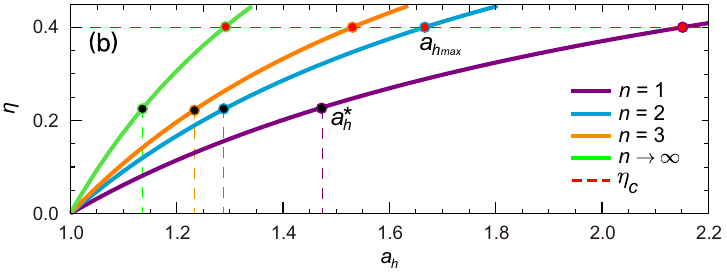}
\caption{\footnotesize{\label{fig3} (a) Work $W^{*}$ vs. potential radius $a_{h}$ curves for each $n = 1, 2, 3 \text{ and } n \rightarrow \infty$ and (b) efficiency $\eta$ vs. potential radius $a_{h}$ together with the Carnot efficiency (red dashed) as comparison. The Figure is the result of an operating cycle in quasi-static state with 60.000 bosons wholly at non-condensed phase. The black dot indicates the efficiency at maximum work for corresponding $n$, whereas the red dot represents the $a_{h_{max}}$ as the maximum radius permitted. Each parameters are $T_{h}=500\text{ nK}, T_{l}=300\text{ nK}, \text{ and } a_{l}=1$}}
\centering
\end{figure}

\begin{figure}[!b]
\includegraphics[width=.24\textwidth]{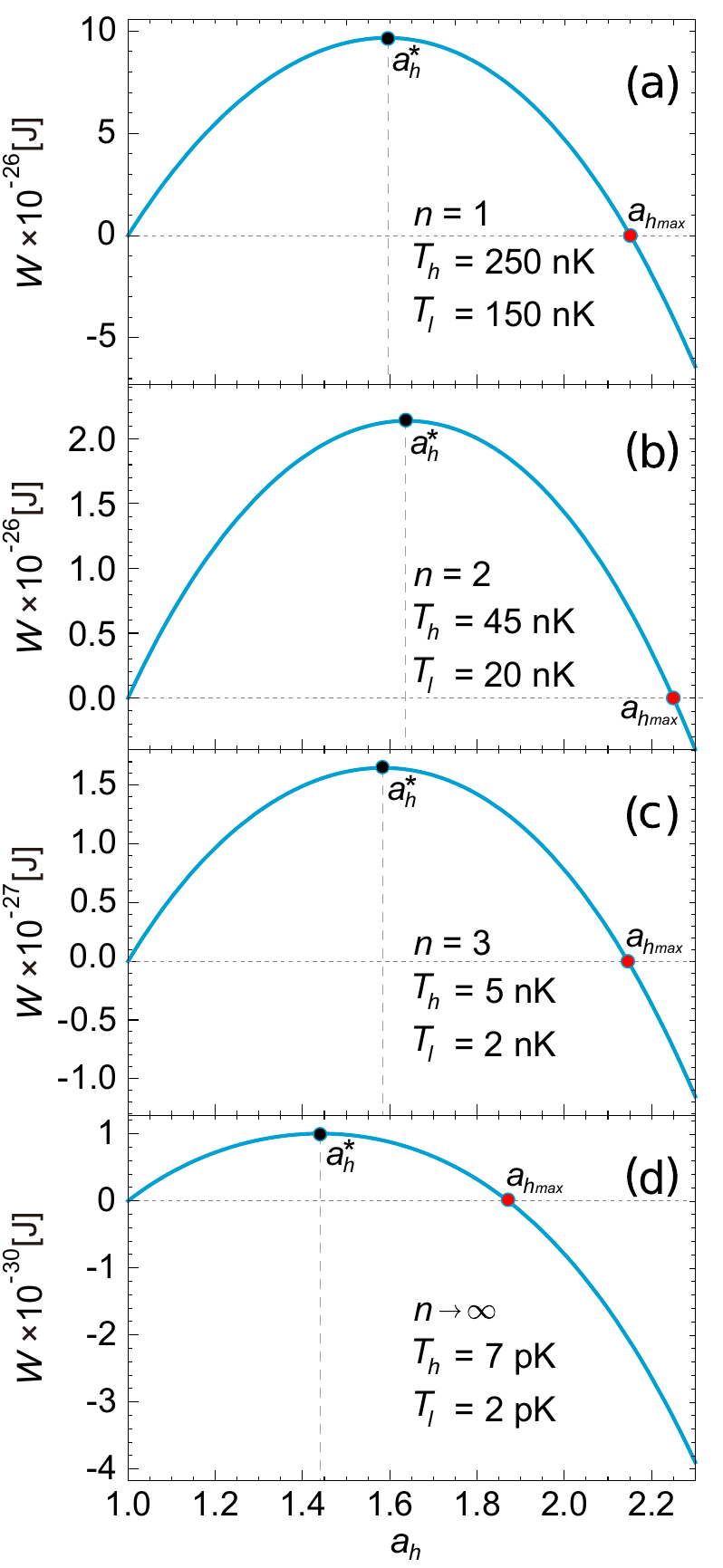}\hfill
\includegraphics[width=.24\textwidth]{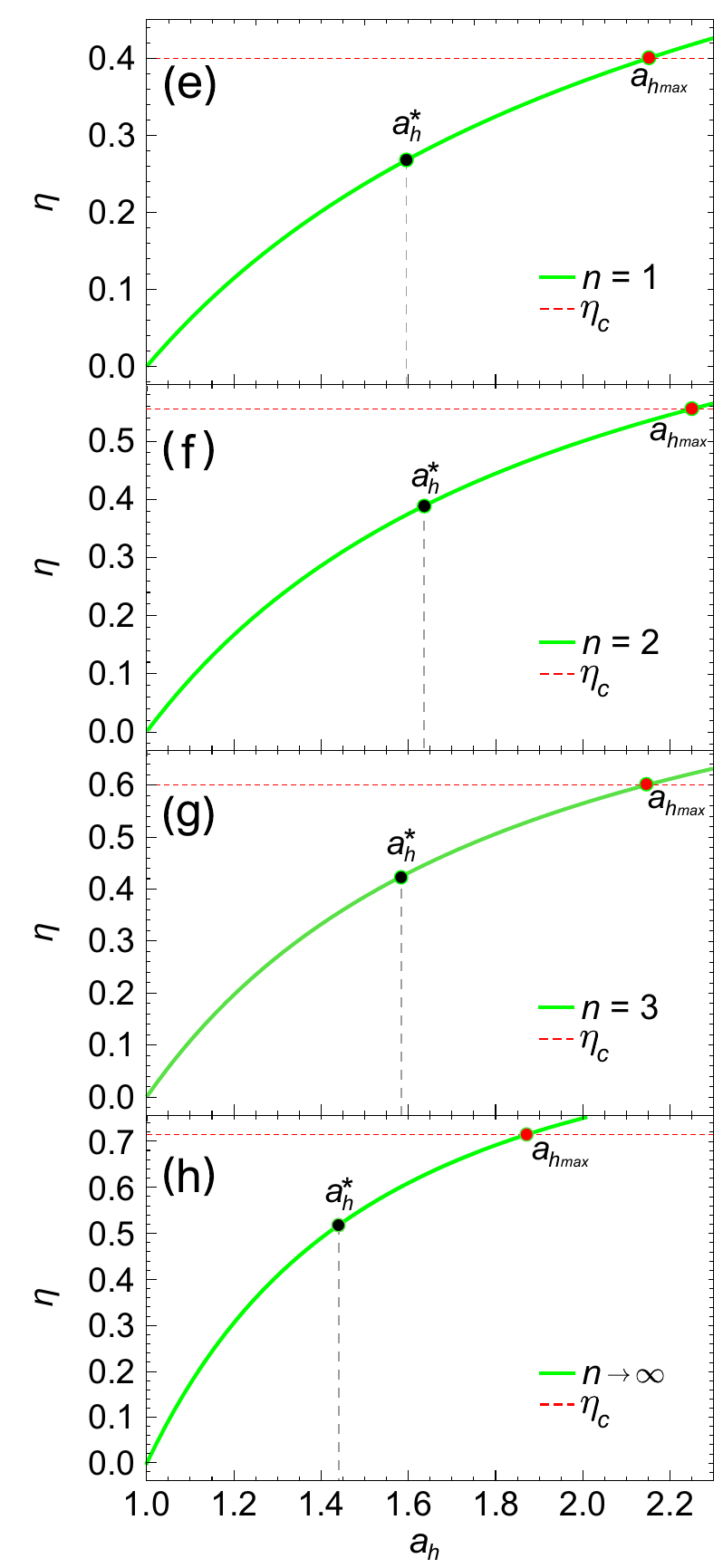}\\
\caption{\footnotesize\label{fig4} The result of an operating cycle in quasi-static manner with 60.000 bosons wholly at condensed phase. The LHS column, (a)$-$(d), shows total work as a function of $a_{h}$, whilst the RHS column, (e)$-$(h), displays the curve of efficiency versus $a_{h}$ along with Carnot efficiency (red-dashed) for each $n = 1, 2, 3$ and $n \rightarrow \infty$ respectively. The black dot and the red dot represent the same meaning as in Figure \ref{fig3}. In this simulation, both work and efficiency use the same $T_{h}$ and $T_{l}$; we set different $T_{h}$ and $T_{l}$ for each $n$ correspond to the $T_{c}$ on Table \ref{tab1}.}
\centering
\end{figure}

In the quasi-static cycle all four-stroke is reversible, as consequences of internal friction due to expansion and compression during isentropic stroke are negligible. Since thermal contact within medium and reservoir last for long time, so we get from Equation \ref{eq16} and \ref{eq21} $T_{3} = T_{h}$ and $T_{1}=T_{l}$. For the cycle in quasi-static regime, we present the result in Figure \ref{fig3} and \ref{fig4}. Figure \ref{fig3} represent the numerical solutions of quasi-static Otto cycle which operated in non-condensed phase or just normal bose gas that is when $T\ge T_{c}$ and Figure \ref{fig4} represent numerical solutions of quasi-static Otto cycle which operated in condensed phase (BEC phase) that is when $T\le T_{c}$. In this simulation parameters $T_{h}$ (hot reservoir temperature), $T_{l}$ (cold reservoir temperature) and $a_{l}$ (initial potential radius) are kept constant, whilst $a_{h}$ (final potential radius) is varied.

Figure \ref{fig3} displays work and efficiency for several values of $n$ for medium in non-condensed phase. Left shows the work curve versus  $a_{h}$, and at each curve the maximum work is marked with a small black dot. For medium in non-condensed phase $T_{l}$ and $T_{h}$ are given equally for all $n$ that is $T_{l} = 300$ nK and $T_{h} = 500$ nK. This is intended so that the resulting work only depends on $n$ so that a comparison can be obtained. Right displays the efficiency curve versus  $a_{h}$ together with Carnot efficiency ($\eta_{C}$) at a constant value of $\eta = 0,4$. $a_{h}$ at maximum work is also shown in this efficiency curve which is denoted by efficiency at maximum work. 

By obtaining the analytical solutions of total work in a cycle, we use Equation \ref{eq12}, which the first line is the solution for condensed phase and the second line is the solution for non-condensed phase. The total work for non-condensed phase is written as
\begin{align}
   W_{t_{ncon}} = \space{}
   &bA'k^{b+1}_{B}a^{3}_{l} \left( 1-\kappa^{\frac{1}{b}} \right) \times \nonumber \\
   &\Big[ T^{b+1}_{h} g_{b+1}(z_{3}) 
   -\kappa^{-\frac{b+1}{b}}T^{b+1}_{l} g_{b+1}(z_{2}) \Big]
   \label{eq26}
\end{align}
$z_{2} \text{ and } z_{3}$ are the fugacity for point 2 and 3 respectively. Since during isentropic stroke there is no heat exchange from medium to the environment, the number of atoms $N$ is fixed \cite{Myers.2022}. As a consequence Equation \ref{eq9} must be constant during this process. By substituting the Temperature and Volume relation from Equation \ref{eq25}, we found fugacity must be constant during isentropic stroke, that is we have $z_{1}(a_{h}, T_{1}) = z_{2}(a_{l}, T_{2})$ and $z_{3}(a_{l},T_{3}) = z_{4}(a_{h},T_{4})$. Furthermore, at high temperature limit, we only need the first term of the expansion of Equation \ref{eq7} because we obtain $\frac{N}{A'a^{3}(k_{B}T)^{b}}\ll 1$ by substituting an appropriate value of $T$. Therefore, Equation \ref{eq26} is transformed to classical formulation of
\begin{align}
   W_{t} = b N k_{B} \left( 1 - \kappa^{\frac{1}{b}} \right) \left[T_{h} - \kappa ^{-\frac{1}{b}}T_{l}\right]
   \label{eq27}
\end{align}
For fixed $T_{h}$ and $T_{l}$ value, the total work decreases as $n$ increases, as shown in Figure \ref{fig3}. These results are in agreement with \cite{Zheng.2014}, which at constant external energy $\epsilon_{0}$ whilst volume is varied, the total work produced in a cycle decreases with increasing degree of potential $n$.

In general, the efficiency of Equation \ref{eq23} can be expressed as
\begin{align}
    \eta = 1 - \kappa^{\frac{1}{b}}
    \label{eq28}
\end{align}
Furthermore, efficiency increases with increasing $n$, from Figure \ref{fig3} we see that for small compression, (small $a_{h}$), engines operating at high $n$ are more efficient than engines operating at small $n$. These results are also consistent with those obtained in references \cite{Zheng.2014}.

Efficiency at maximum work is equal for all $n$, which is exactly Curzon-Ahlborn efficiency \cite{CA.1975}. By deriving equation \ref{eq27} with respect to $\kappa$, we get $\kappa^{\frac{1}{b}}_{max}=\left(T_{l}/T_{h}\right)^{\frac{1}{2}}$ so $\eta = 1-\left(T_{l}/T_{h}\right)^{\frac{1}{2}}$. On Section  \ref{sec4.2}, we also see that efficiency at maximum power for medium in non-condensed phase is also equivalent to Curzon-Ahlborn efficiency.

In Figure \ref{fig4} we represent the total work for several $n$ for medium in condensed phase. $T_{l}$ and $T_{h}$ are chosen in such a way for each $n$ medium is always below critical temperature. Need to be noted that for each $n$, we cannot display them in one figure as we did in Figure \ref{fig3}, this is due to the difference in critical temperature at each $n$. We get the analytical expression for total work for medium in condensed phase below
\begin{align}
   W_{t_{con}} = \space{}
   &bA'k^{b+1}_{B}a^{3}_{l} \left( 1-\kappa^{\frac{1}{b}} \right) \zeta(b+1) \times \nonumber \\ 
   &\left[ T^{b+1}_{h}-\kappa^{-\frac{b+1}{b}}T^{b+1}_{l} \right]
   \label{eq29}
\end{align}
$\zeta(b+1)$ is a zeta function originated from Equation \ref{eq7}, i.e. when $z=1$. By dint of choosing the decreasing temperature as $n$ increases, the maximum work obtained is also decreased as $n$ increases as shown in Figure \ref{fig4}. But unlike in the non-condensed phase, where the optimum efficiency is the same for all n, the efficiency at the maximum work increases with increasing n. It shows that the optimum efficiency for medium in condensed phase is dependent on the properties of the medium. 

In Section \ref{sec4.2}, it is clearly seen that the efficiency at maximum power for medium in condensed phase is higher than Curzon-Ahlborn efficiency.  Need to be noted that work in condensed phase is not explicitly a function of $N$ (number of particles) because the number of condensed bosons is a function of temperature. Referring to Equation \ref{eq9}, we obtain the fraction of excited boson to total $N$ ($T \le T_{c}$) as written
\begin{align}
    \frac{N_{T}}{N} = \left( \frac{T}{T_{c}} \right)^{b}
    \label{eq30}
\end{align}
The maximum excitation occurs when $T=T_{c}$ whilst minimum excitation occurs when all bosons are condensed, viz. when $T=0$.

To obtain the ideal compression ratio (W is maximized), the Equation \ref{eq29} is derived by $\kappa$
\begin{align}
    T^{b+1}_{l} \left[ (b+1) - (b\kappa^{*})^{\frac{1}{b}} \right] - T^{b+1}_{h} (\kappa^{*})^{\frac{2+b}{b}} = 0
    \label{eq31}
\end{align}
$\kappa^{*} \text{ and } a^{*}$ are compression ratio and radius of potential at maximum work, respectively. The ideal  $\kappa^{*} \text{ and } a^{*}$ for any number of $n$ can be derived by solving the Equation \ref{eq31} which the values are displayed in table \ref{tab2}. 

\begin{table}[htp]
\begin{ruledtabular}
\caption{\footnotesize The engine parameters at non-condensed and condensed phase according to Figure \ref{fig3} and \ref{fig4}, respectively.} 
\label{tab2}%
\begin{tabular}{llllll}
 \multirow{2}{*}{$n$}
   &\multicolumn{2}{l}{$non-condensed$}
   &\multicolumn{3}{c}{$condensed$}\\
   {} &$a^{*}_{h}$ &$a_{h_{max}}$ &$a^{*}_{h}$ &$a_{h_{max}}$ &$\kappa^{*}$\\
   \hline
1         &1,47   &2,17   &1,59   &2,16   &0,24\\
2         &1,29   &1,65   &1,63   &2,25   &0,22\\
3         &1,23   &1,53   &1,58   &2,15   &0,25\\
$\infty$  &1,13   &1,28   &1,44   &1,82   &0,33\\
\end{tabular}
\end{ruledtabular}
\end{table}

Based on the Table, the values do not change constantly as the variation of $n$ is increased. It is due to the total work done in the shape of a sphere ($n = 2$) being equally distributed. Thus, the compression ratio required is lesser than other shapes. Because we are not really interested in engine that operates quasi-statically, due to long and infinite stroke time the power it produces is small to zero. We are more interested in finite time Otto cycle. The contact between medium and reservoir during the isochoric stroke is finite,  medium is not in equilibrium with the thermal reservoir. But still during expansion and compression isentropic stroke are slow and quasi-static.

\subsection{Endoreversible Performance} \label{sec4.2}
First of all, we review the efficiency and power output at condensed phase when the medium is under critical temperature. By substituting Equation \ref{eq13}, \ref{eq17}, and \ref{eq21} to Equation \ref{eq23}, we derived similar efficiency as the quasi-static condition:
\begin{align}
    \eta_{T \le T_{C}} = 1 - \kappa^{\frac{1}{b}}
    \label{eq32}
\end{align}
while the power output is derived by combining Equation \ref{eq17} and \ref{eq22} with Equation \ref{eq24}, so as by using the relation of temperature in Equation \ref{eq16}, \ref{eq21}, and \ref{eq25}. As a result, the power output of condensed phase can be written below 

\begin{figure*}[!htp]
\includegraphics[scale=0.60]{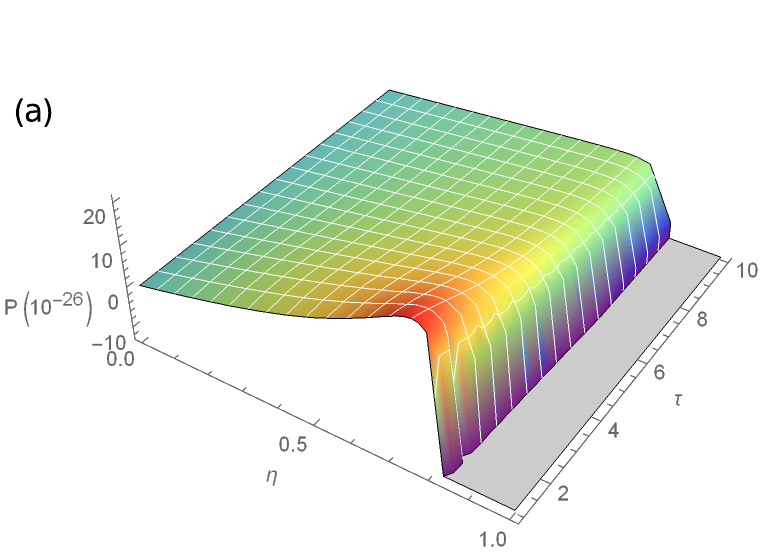} \hfill
\includegraphics[scale=0.35]{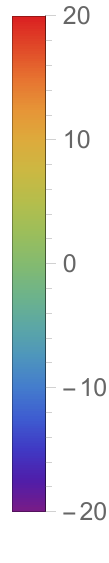} \hfill
\includegraphics[scale=0.60]{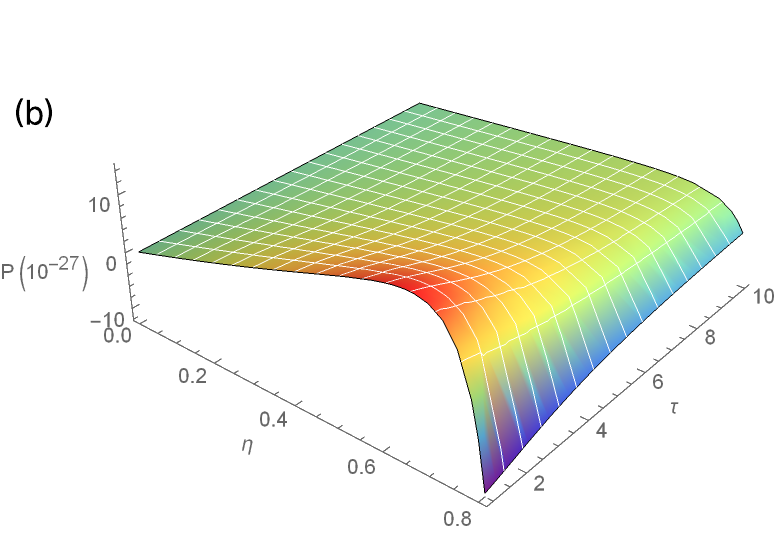} \hfill
\includegraphics[scale=0.35]{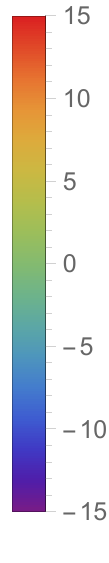} \hfill
\includegraphics[scale=0.60]{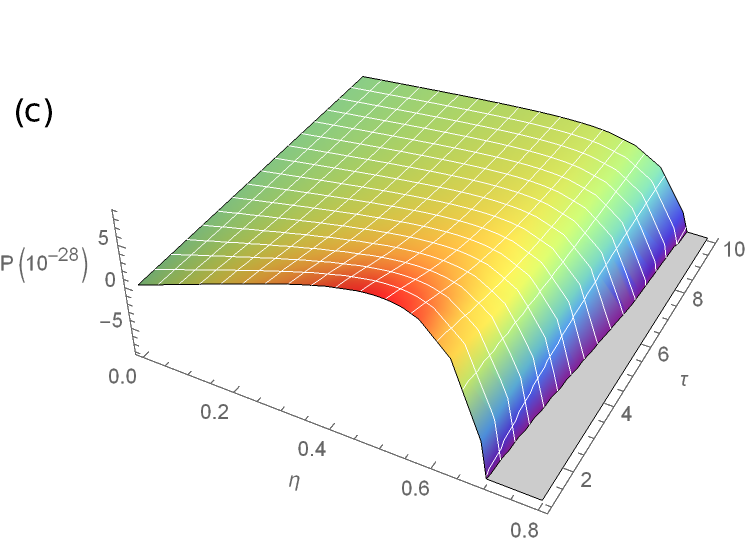} \hfill
\includegraphics[scale=0.35]{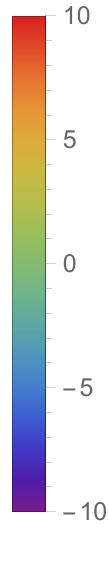} \hfill
\includegraphics[scale=0.60]{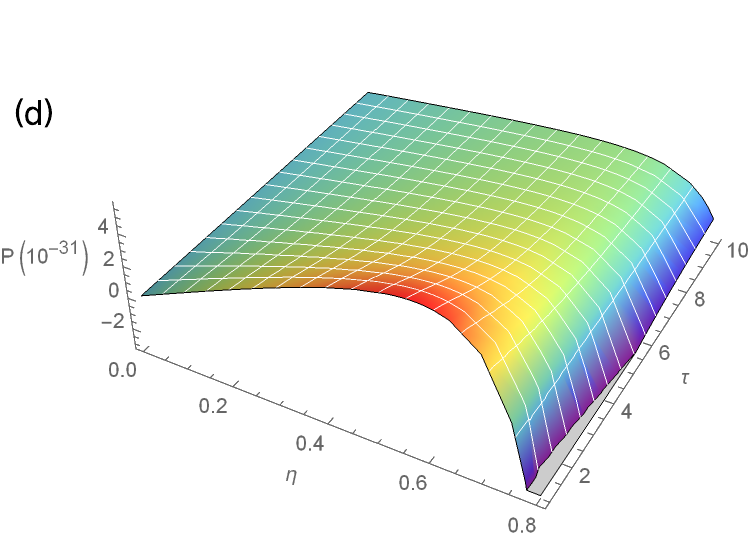}\hfill
\includegraphics[scale=0.41]{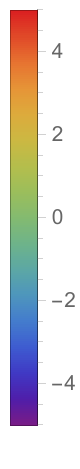} \hfill
\caption{\footnotesize\label{fig5} 3D plot of power as function of $\eta$ and $\tau$ for medium in  BEC phase for (a) $n = 1$, (b) $n = 2$, (c) $n = 3$, and (d) $n \rightarrow \infty$, with hot and cold reservoir temperatures used are $T_{h}=300$ nK and $T_{l}=50$ nK, $T_{h}=45$ nK and $T_{l}=10$ nK, $T_{h}=6$ nK and $T_{l}=2$ nK, as well as $T_{h}=8$ pK and $T_{l}=2$ pK, respectively. The parameters are $\alpha_{l} = \alpha_{h} = \gamma = 1$.}
\centering
\end{figure*}

\begin{widetext}
\begin{align}
    \scalemath{0.85}{P_{T \le T_{C}} = \frac{bA'k^{b+1}_{B} g_{b+1}(1)a^{3}_{l} \left(1 - \kappa^{\frac{1}{b}}\right) \left[\left[ T_{h} \kappa^{\frac{1}{b}} e^{\alpha_{l}\tau_{l}} \left( e^{\alpha_{h}\tau_{h}}-1 \right) + T_{l} \left( e^{\alpha_{l}\tau_{l}}-1 \right) \right]^{b+1} - \left[ T_{l} e^{\alpha_{h}\tau_{h}} \left( e^{\alpha_{l}\tau_{l}}-1 \right) + T_{h} \kappa^{\frac{1}{b}} \left( e^{\alpha_{h}\tau_{h}}-1 \right) \right]^{b+1} \right]}{\gamma (\tau_{c}+\tau_{h})\kappa^{\frac{b+1}{b}} \left( e^{\alpha_{h}\tau_{h}+\alpha_{c}\tau_{c}}-1 \right)^{b+1}}}
    \label{eq33}
\\ \nonumber
\end{align}
\end{widetext}
which each $T_{h}$ and $T_{l}$ are the hot and cold reservoir temperatures, respectively, while each $\alpha_{h}$ and $\alpha_{l}$ represent the thermal conductivity while making contact with hot and cold reservoir, and the time of strokes within the heating and cooling process are denoted with $\tau_{h}$ and $\tau_{l}$.

\begin{figure*}[!htp]
\includegraphics[scale=0.65]{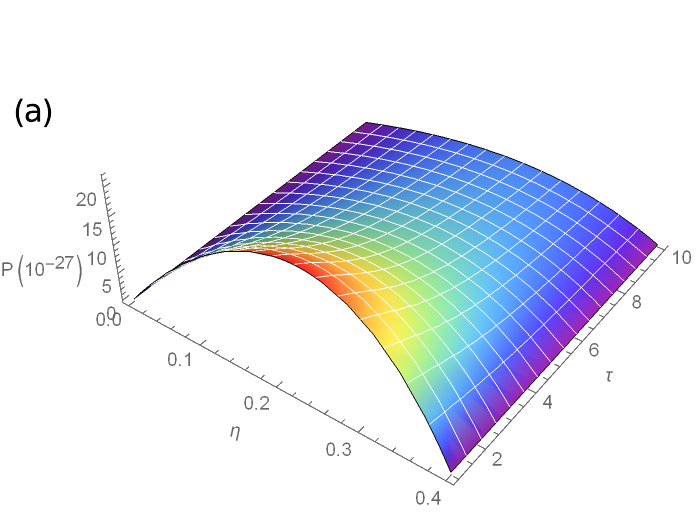} \hfill
\includegraphics[scale=0.45]{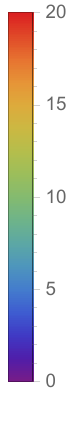} \hfill
\includegraphics[scale=0.65]{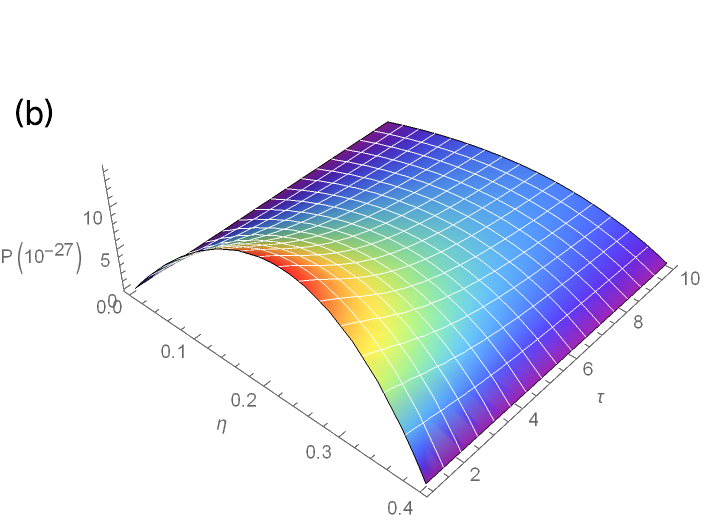} \hfill
\includegraphics[scale=0.45]{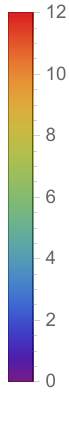} \hfill
\includegraphics[scale=0.65]{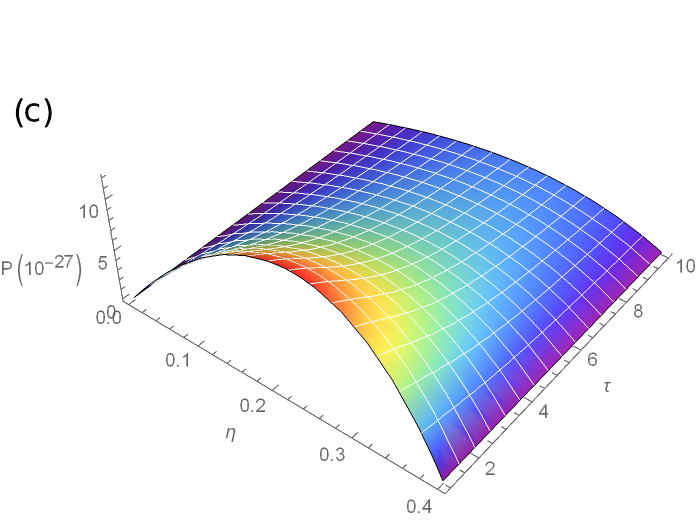} \hfill
\includegraphics[scale=0.45]{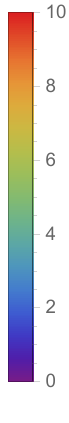} \hfill
\includegraphics[scale=0.65]{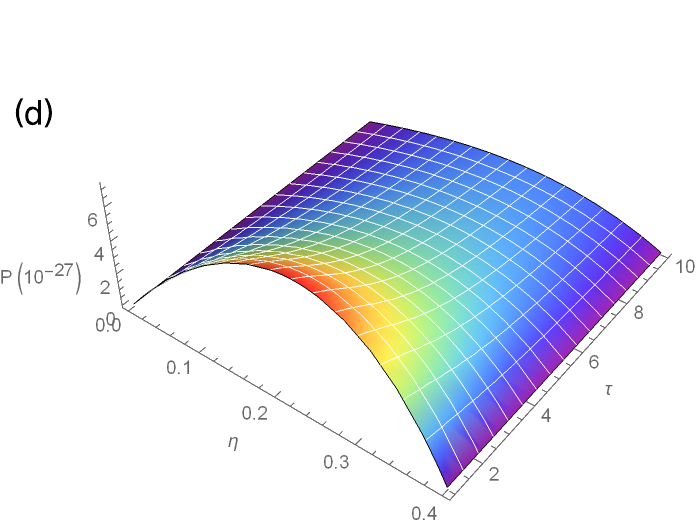}\hfill
\includegraphics[scale=0.45]{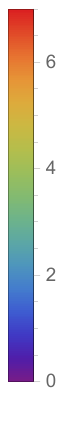} \hfill
\caption{\footnotesize\label{fig6} 3D plot of power as function of $\eta$ and $\tau$ for medium in non-condensed phase for (a) $n = 1$, (b) $n = 2$, (c) $n = 3$, and (d) $n \rightarrow \infty$, respectively. Cold and hot reservoir temperatures are 300 nK and 500 nK for all $n$. Other parameters are $\alpha_{l} = \alpha_{h} = \gamma = 1$.}
\centering
\end{figure*}

Furthermore, we obtain the efficiency of the non-condensed phase is the same as condensed phase, 
i.e.
\begin{align}
    \eta_{T \ge T_{C}} = 1 - \kappa^{\frac{1}{b}}.
    \label{eq34}
\end{align}
Since efficiency is determined by the rate of work done during expansion and compression, for endoreversible cycle expansion and compression stroke is operated by isentropic condition means its quasistatic, so there is no internal friction arises which will reduce efficiency \cite{Cakmak2017}. However, the power output will be nonzero since the finite time process at isochoric stroke. This result indicates that in both quasi-static and endoreversible cases has no impact on engine efficiency but does on power. As obtained in the formulation of quasi-static work (Equation \ref{eq26}), the fugacity of Equation \ref{eq7} is approximated only at the first term because the $z$ value is small. $T_{3}$ and $T_{2}$ is derived from Equation \ref{eq16}, \ref{eq21}, and \ref{eq25} so that the obtained power output is explicitly depend on $N$
\begin{widetext}
\begin{align}
    \scalemath{0.85}{P_{T \ge T_{C}} = \frac{bNk_{B} \left(1 - \kappa^{\frac{1}{b}}\right) \left[\left[ T_{h} \kappa^{\frac{1}{b}} e^{\alpha_{l}\tau_{l}} \left( e^{\alpha_{h}\tau_{h}}-1 \right) + T_{l} \left( e^{\alpha_{l}\tau_{l}}-1 \right) \right] - \left[ T_{l} e^{\alpha_{h}\tau_{h}} \left( e^{\alpha_{l}\tau_{l}}-1 \right) + T_{h} \kappa^{\frac{1}{b}} \left( e^{\alpha_{h}\tau_{h}}-1 \right) \right] \right]}{\gamma (\tau_{c}+\tau_{h})\kappa^{\frac{1}{b}} \left( e^{\alpha_{h}\tau_{h}+\alpha_{c}\tau_{c}}-1 \right)}}
    \label{eq35}
\end{align}
\end{widetext}
need to be noted, at $T \ge T_{c}$, all bosons are in non-condensed phase and there is no condensed boson yet. In contrast, at $T \le T_{c}$, the amount of condensed boson depends on temperature \ref{eq30} so that $N$ can not be written explicitly. By replacing $\kappa^{\frac{1}{b}}$ with $1-\eta$, power can also be represented as a function of efficiency. We visualize power as a function of efficiency ($\eta$) and isochoric stroke time ($\tau$) by a 3D plot in Figure \ref{fig5} for medium in BEC phase and in Figure \ref{fig6} for medium in non-BEC phase.

We found that longer stroke times minimize power. This is due to the dependence of the denominator of the equation \ref{eq33} on stroke time. Power is also minimized as efficiency approaches 1; it shows that engine performance is not only seen from its efficiency but also from the amount of power produced. By the reason of that, it is interesting to know at which efficiency the power becomes maximum. Efficiency at maximum power (EMP) is marked with a peak on each curve. The apex of this curve shifts to the left as $\tau$ increases, which means that EMP also decreases as $\tau$  increases. However, at high $n$ the peak shift of the curve is small, so the increasing of $\tau$ does not have a significant effect on the decreasing of EMP.

\begin{figure}[!htp]
\includegraphics[width=0.49\textwidth]{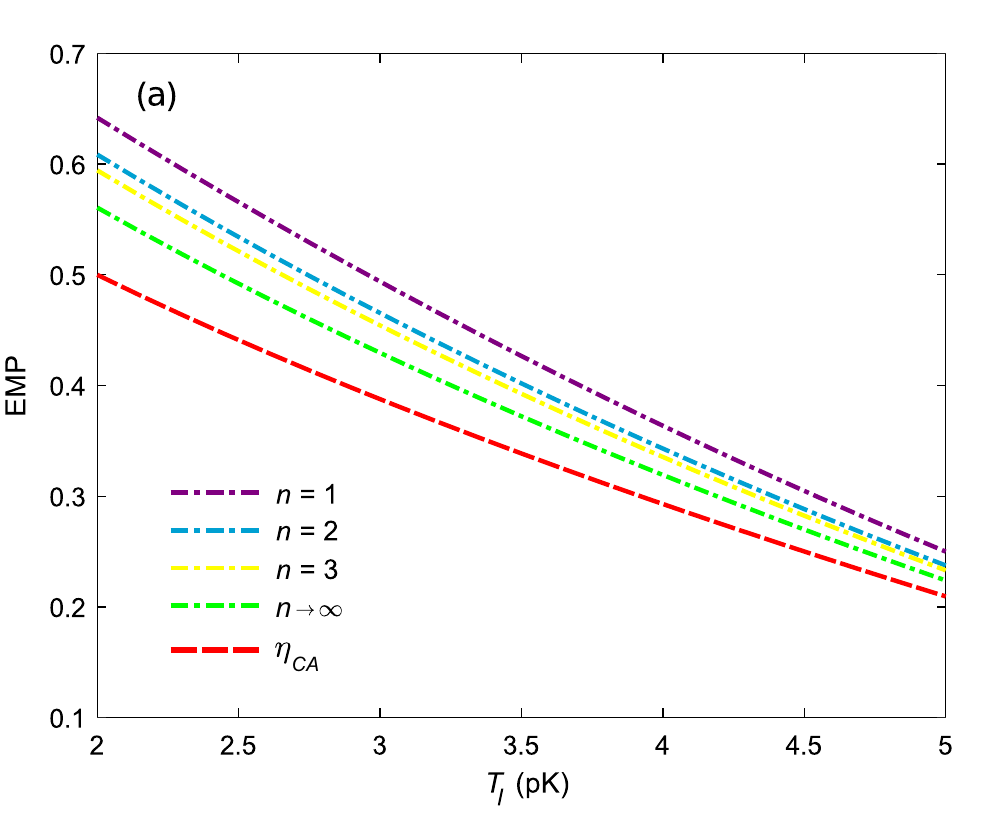} \\
\includegraphics[width=0.235\textwidth]{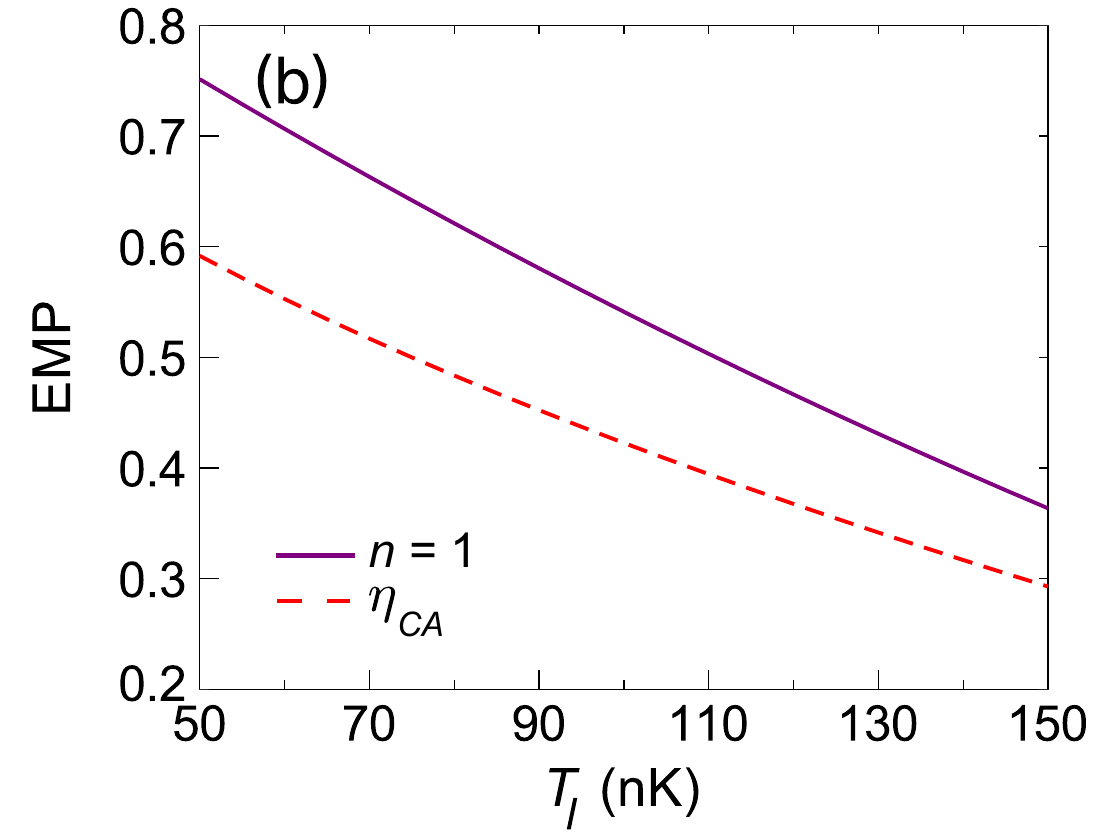} 
\includegraphics[width=0.235\textwidth]{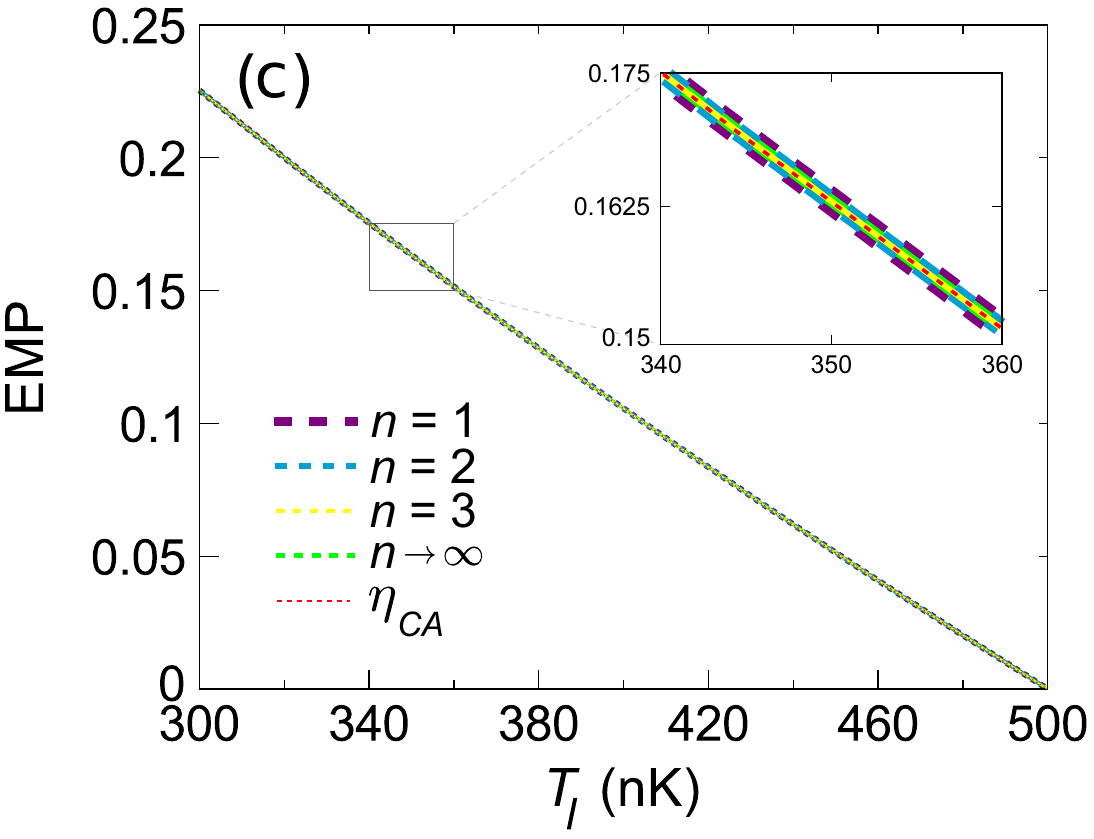}
\\
\caption{\footnotesize\label{fig7} (a) EMP vs $T_{l}$ for medium in BEC phase for all $n$, using $T_{h} = 8$ pK. (b) EMP vs $T_{l}$ for medium in condensed phase for $n=1$, using $T_{h} = 300$ nK. (c) EMP for medium in non-condensed phase for all $n$, using $T_{h} = 500$ nK. On each curve, Curzon-Ahlborn efficiency (dashed red line) is given as a comparison. Others parameter are $\tau_{l} = \tau_{h} = \alpha_{l} = \alpha_{h} = \gamma = 1$.}
\centering
\end{figure}

We used $T_{h}$ and $T_{l}$ in this non-condensed phase slightly higher than the critical temperature of $n=1$, so that none of the $n$ has condensed. We also use this value in representing EMP in Figure \ref{fig7}. Although the shape of the curves are similar for all $n$, the power $P$ decreases with the increasing of $n$, which is also confirmed by Figure \ref{fig3}a. Unlike the power in the BEC phase where the peak of the curve shifts slightly with an increasing of $\tau$, in non-condensed phase, the peak of the curve does not change with increasing of $\tau$. This shows that the efficiency at maximum power of non-condensed phase does not depend on $\tau$. As has been found in the quasi-static results, the efficiency at maximum power for the medium in the non-condensed phase is the Curzon-Ahlborn efficiency which only depends on the temperature of the reservoir. Curzon-Ahlborn efficiency is given by $\eta = 1-\left(T_{l}/T_{h}\right)^{\frac{1}{2}}$ so we get $\eta=0.22$ as exactly shown in Figure \ref{fig6} The optimum efficiency achieved in the non-condensed phase is also lower than in the BEC phase. 

It is well established that there is an inherent trade-off between efficiency and power \cite{BCakmak.2019}. Efficiency is maximized at long and infinite stroke time, but power is minimized for long stroke time due to dependence of the denominator in Equations \ref{eq33} and \ref{eq35}. The highest efficiency is bounded by Carnot efficiency, that is when $\kappa^{\frac{1}{b}} = T_{l}/T_{h}$. By substituting this to Equation \ref{eq33} or \ref{eq35}, we found that both in condensed phase and in non-condensed phase, power vanishes at that value. So it is important to find an optimum efficiency, Efficiency at Maximum Power (EMP). Similar to what Curzon and Ahlborn \cite{CA.1975} did, then followed by many researchers \cite{Myers.2022, Deffner.2018, Wang.2009, MyersDeffner.2020, WangJ.2012, WangR.2013, ZSmith.2020}, we also determine EMP by maximizing Equations \ref{eq33} and \ref{eq35} with respect to $\kappa$, then the results are substituted back to Equations \ref{eq32} and \ref{eq34}, which shown in Figure \ref{fig7}.

Since the critical temperature of $n \rightarrow \infty$ is the lowest, we used $T_{h}$ and $T_{l}$ which are comparable to the critical temperature of $n \rightarrow \infty$, thus all $n$ is in the condensation phase. As shown in Figure \ref{fig7}a, EMP decreases as increasing of $n$ and temperature of cold reservoir $T_{l}$, but still higher than Curzon-Ahlborn efficiency. Because of work in a quantum system is manifested by the difference between the initial and final energies of the system \cite{roncaglia2014quantum}, this result can be linked to the change in internal energy during expansion and compression. Based on equation \ref{eq12}, the internal energy depends on the order of $(1/n)$, so $n=1$ gives the highest expression of internal energy, and it reduces with the increasing of $n$. The power and EMP are inversely proportional to $n$. Moreover, EMP also shifts to Curzon-Ahlborn efficiency as the temperature of cold reservoir $T_{l}$ increases, which can be seen in Figure \ref{fig7}b and Figure \ref{fig7}c. 

Furthermore, although at $n = 1$ the highest EMP is produced, the power generated at picokelvin range is the lowest because $P$ depends on the power of $b+1$ (see Equation \ref{eq12}). In addition, $n = 1$  has the largest $b$ while $T \ll 1$, so the power generated is also very small. Therefore, we present the EMP for $n = 1$ separately in Figure \ref{fig7}b in nanokelvin range. In that case, we use $T_{h} = 300$ nK, slightly higher than its critical temperature, which resulting much higher power than in picokelvin range. Classically, work is produced from the movement of the piston when pressure is applied \cite{Myers.2022}, but the atoms in the condensation phase cannot accept or use the pressure exerted on it \cite{Pathria.2011}. Nevertheless, \cite{Myers.2022} claimed that work must be generated by the fraction of bosons out of condensate in the thermal cloud. At $n=1$, the picokelvin temperature is very far below its critical temperature (see Equation \ref{eq30}), and almost all of the bosons have condensed, which causes only a few to be in the thermal cloud. Meanwhile, in the nanokelvin temperature range, which is slightly below its critical temperature, bosons are just starting to condense which causes some of them to still be in the thermal cloud. For this reason, the work and power generated in the nanokelvin range will be greater than the picokelvin range; this is as shown in Figure \ref{fig7}, the EMP in Figure \ref{fig7}b is higher than in Figure \ref{fig7}a. 

\begin{figure*}[htp!]
\centering
\includegraphics[width=0.96\linewidth]{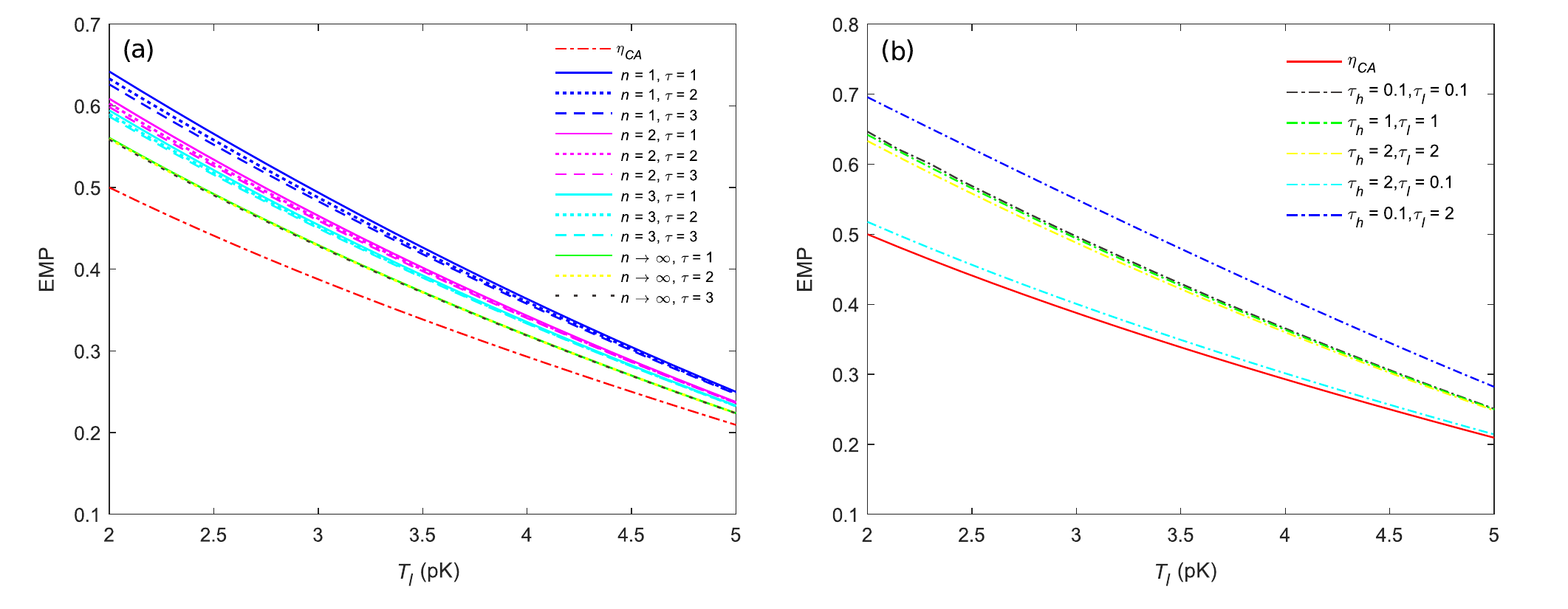}
\caption{\footnotesize\label{fig8} EMP at BEC phase with variation in isochoric stroke time. (a) For all $n$ with heating and cooling time equated, the solid line represents $\tau = 1$, dotted line represents $\tau = 2$, and dashed line represents $\tau = 3$. (b) For only $n=1$ with heating and cooling stroke time is differentiated. For comparison in each curve Curzon-Ahlborn efficiency $(\eta_{CA})$ is also given. In both simulation we consider $T_{h} = 8$ pK while other parameters are $\alpha_{l} = \alpha_{h} = \gamma = 1$.}
\end{figure*}

At non-condensed phase, we obtained the optimal compression ratio as $\kappa_{max}=(T_{l}/T_{h})^{b/2}$. By substituting it into Equation \ref{eq34}, we get $\eta=1-(T_{l}/T_{h})^{1/2}$ in which is Curzon-Ahlborn efficiency. As shown in Figure \ref{fig7}c, EMP has the same value for all $n$. As we did previously, at high temperature limit, we took $\frac{N}{A'a^{3}(k_{B}T)^{b}}\ll 1$ where the term $\frac{N}{A'a^{3}}$ is the quantum property of bosons in a certain potential with the dimensions of energy. If $(k_{B}T)^{b}$ is much greater than $\frac{N}{A'a^{3}}$, then the quantum property of the gas can be neglected. Therefore, the energy of the bosons which was initially discrete has become continuous. When the energy of system matches classical conditions, the efficiency at maximum power is equal to the Curzon-Ahlborn efficiency \cite{Kosloff.2017}. However, when $\frac{N}{A'a^{3}}$ is comparable or greater than $(k_{B}T)^{b}$, the quantum effect of the gas cannot be neglected. Hence, the EMP also depends on a quantum property of the gas, including potential. Each potential gives different energy eigenvalue \cite{mahajan2020quantum}, so that it will affect the performance of the engine. Not only the difference in energy quantization affects performance, but in this study, we also found that the variations in the thermal contact time with the reservoir during isochoric strokes also affect the performance, especially the EMP. We visualize EMP for working medium in BEC phase as a function of cold bath temperature with different heating and cooling stroke time, $\tau_{h}$ and $\tau_{l}$ respectively. In this case, we only represented EMP in BEC phase since in non-condensation phase EMP is independent of heating and cooling stroke time. 

In Figure \ref{fig8}a, we consider the engine works in an equal heating and cooling stroke time, $\tau_{l} = \tau_{h}$, which are varied in time units for all $n$. On the other hand, in Figure \ref{fig8}b, the heating and cooling stroke time are distinguished for only $n=1$, the highest EMP producer, whereas all other parameters are kept constant. As seen in Figure \ref{fig8}a, the increasing of stroke time relatively decreases EMP even though at higher $n$ does not give a significant difference. This result is not the same as obtained in prior research \cite{Myers.2022} in which the increasing of stroke time precisely increases the EMP produced. Ideally, the longer the stroke time makes the engine more likely reversible, and the engine that can exhibit reversibility would provide the highest efficiency \cite{Cengel.2008}. However, there are also some interesting things we found in this study. Figure \ref{fig8}b shows that the EMP produced is significantly higher at short heating stroke time and long cooling stroke time. Otherwise, at long heating stroke time and a short cooling stroke time, the EMP produced is lower than other configurations even so close to Curzon-Ahlborn efficiency. Furthermore, EMP at $\tau_{h}<\tau_{l}$ is higher than EMP at $\tau_{h}=\tau_{l}$. Physically, it is due to the medium extracting more energy from hot reservoir when the heating stroke time is long, so that the temperature of medium reaches the temperature of hot reservoir itself. In contrast, during the short heating and long cooling stroke time, the medium ejects more energy into the cold reservoir so that the temperature drops. This is in agreement with Figure \ref{fig7} that EMP is high at lower temperature and likewise. 

The increasing efficiency at short and finite heating stroke time is known as the effect of residual coherence due to incomplete thermalizations \cite{PhysRevE.103.032144}. In order to obtain a higher amount of power output, the cycle time should be cut down. At least there are two ways we can do this; first, by cutting the time during the expansion and compression strokes (shortcut to adiabaticity) \cite{Campo2014, OAbah.2019, Cakmak2017, PhysRevLett.111.100502}, and second by cutting the contact thermal time between the medium and the reservoir \cite{PhysRevE.103.032144, PhysRevA.99.062103, PhysRevLett.113.260601}. Nevertheless, in this study, we use an endoreversible cycle with the expansion and compression processes done in isentropic, i.e. quasi-static, so that we can only cut the time from isochoric processes. Furthermore, due to rapid compression and expansion, internal friction (quantum friction) arises and directly reduces efficiency \cite{Cakmak2017}. Finite-time transformation also exhibits entropy production, which leads the engine to irreversibility \cite{Cengel.2008}. However, this irreversibility does not only occur during expansion and compression strokes but also in the heating and cooling strokes. Due to the short heating and cooling stroke time, the medium never reaches thermal equilibrium with the reservoir, thereby leaving the medium in a coherent state in the energy basis $|E_{n}\rangle$, or it is known as residual coherence. The coherence in this energy basis can be associated with entropy production and quantum friction \cite{PhysRevA.99.062103}. Longer heating stroke time increases entropy production because it makes more heat flow from the hot reservoir to the medium. Mathematically, entropy is the amount of heat that flows for the increasing of every one degree in temperature \cite{schroeder1999introduction}, so that the longer the contact time of the medium to the hot reservoir, the higher entropy increases. The flowing heat does not occur continuously due to a thermal equilibrium will be achieved at a certain temperature that the entropy will not increase anymore.

\begin{figure*}[!htbp]
\begin{tabular}{ccccc}
\multicolumn{2}{c}{$n = 1$} &{} &\multicolumn{2}{c}{$n = 2$} \\ 
$\Delta S^{*}_{heating}$ &$\Delta S^{*}_{cooling}$ & &$\Delta S^{*}_{heating}$ &$\Delta S^{*}_{cooling}$ \\
\noalign{\smallskip}
\vspace{0.2cm}
\rotatebox[origin=c]{90}{\scriptsize{$\tau_h = [0, 10]$; $\tau_l = [0, 1]$}}
\includegraphics[scale=0.25, valign=m]{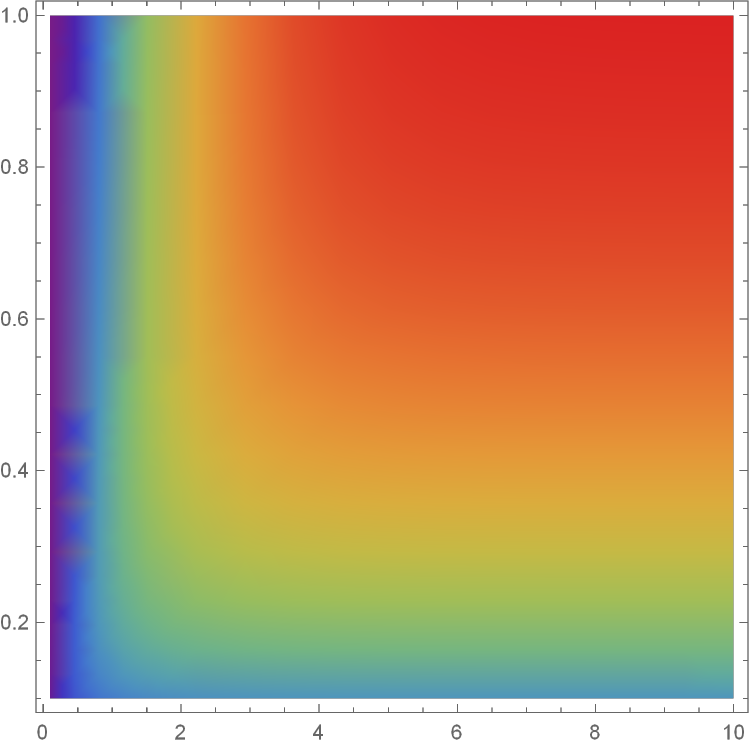} \hfill
\includegraphics[scale=0.30, valign=m]{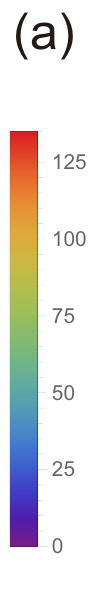} \hfill
&\includegraphics[scale=0.25, valign=m]{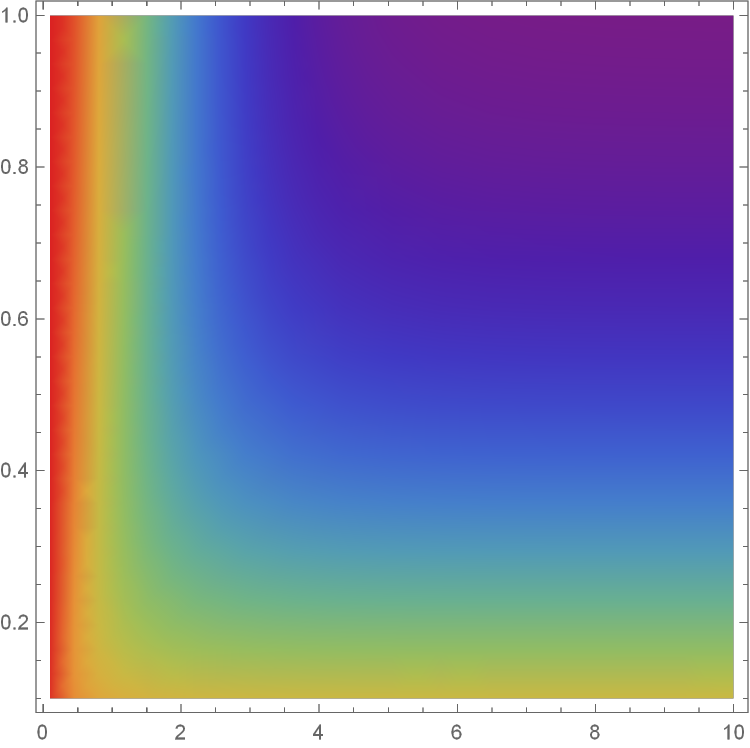} \hfill
\includegraphics[scale=0.30, valign=m]{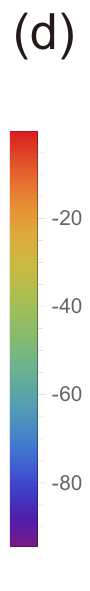} \hfill
&{}
&\includegraphics[scale=0.25, valign=m]{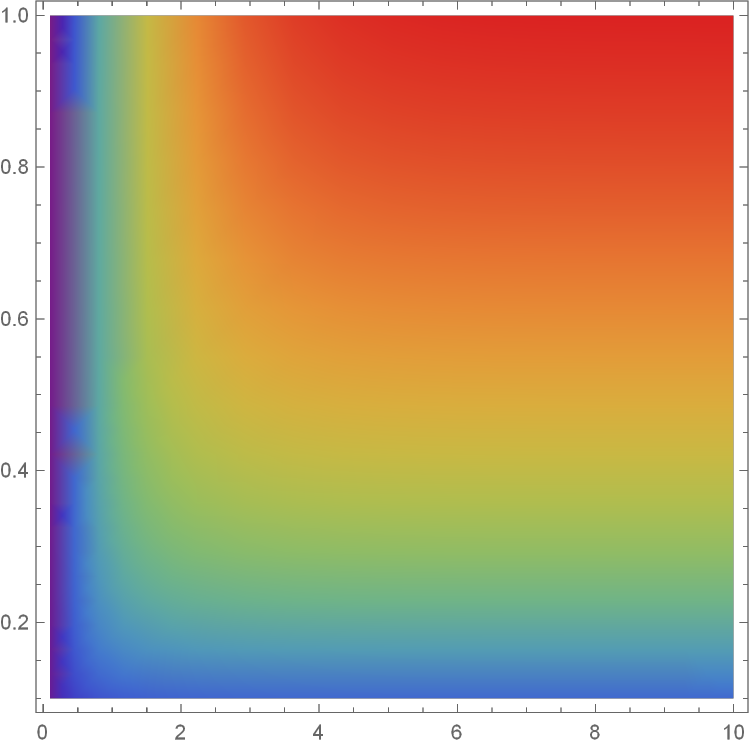} \hfill
\includegraphics[scale=0.30, valign=m]{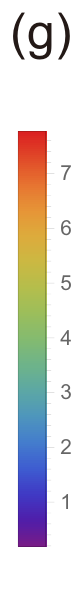} \hfill
&\includegraphics[scale=0.25, valign=m]{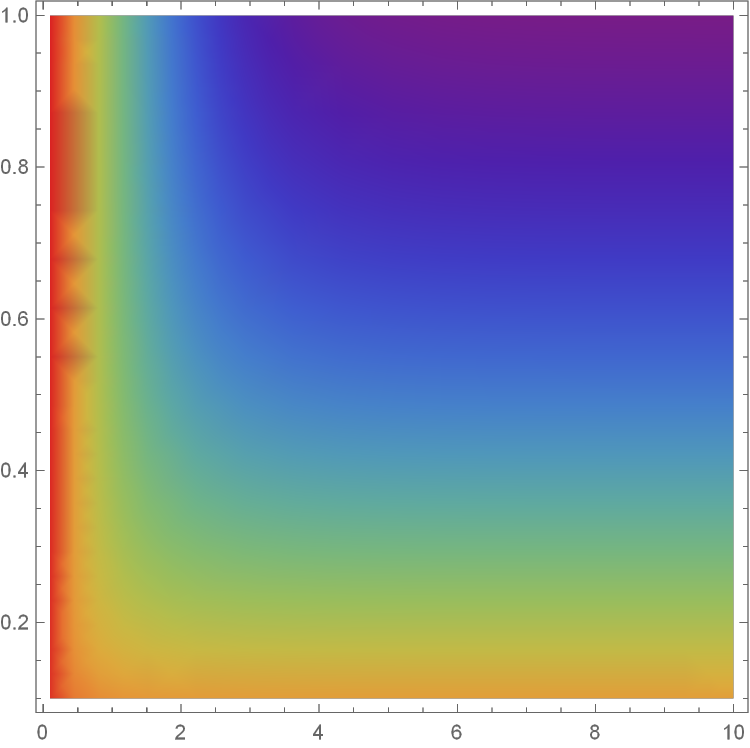} \hfill
\includegraphics[scale=0.30, valign=m]{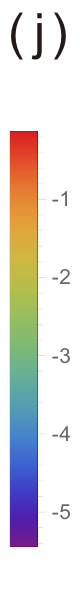} \\

%\noalign{\smallskip}
\vspace{0.2cm}
\rotatebox[origin=c]{90}{\scriptsize{$\tau_h = [0, 1]$; $\tau_l = [0, 1]$}}
\includegraphics[scale=0.25, valign=m]{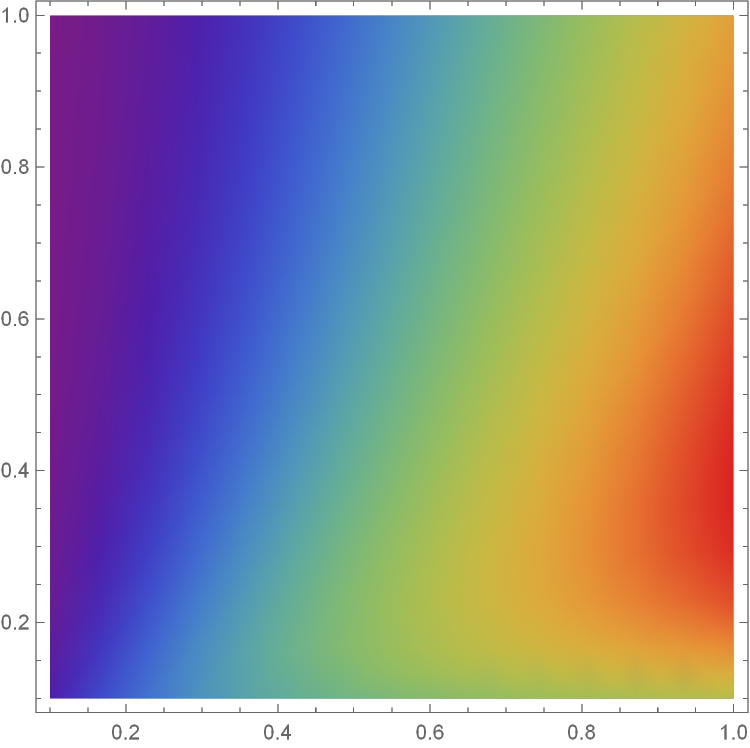} \hfill
\includegraphics[scale=0.30, valign=m]{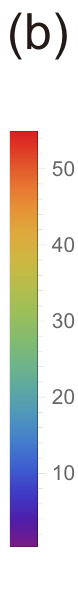} \hfill
&\includegraphics[scale=0.25, valign=m]{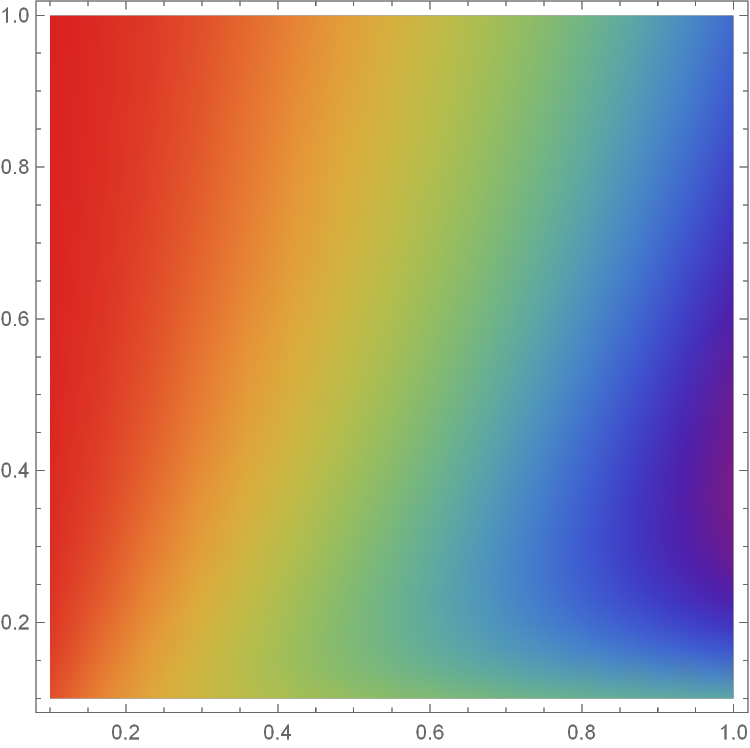} \hfill
\includegraphics[scale=0.30, valign=m]{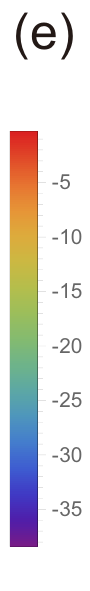} \hfill
&{}
&\includegraphics[scale=0.25, valign=m]{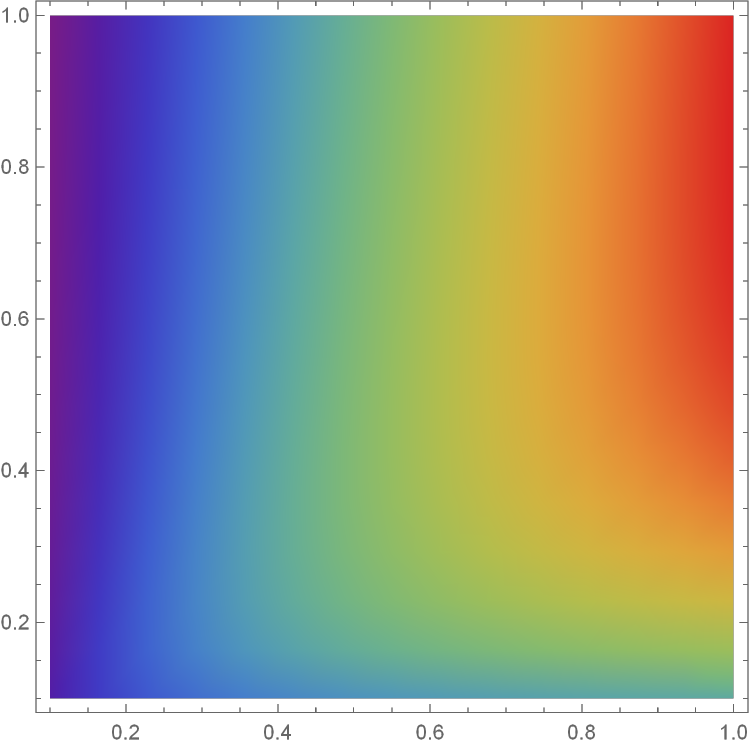} \hfill
\includegraphics[scale=0.30, valign=m]{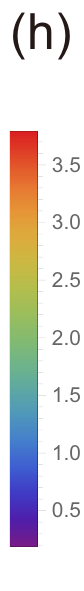} \hfill
&\includegraphics[scale=0.25, valign=m]{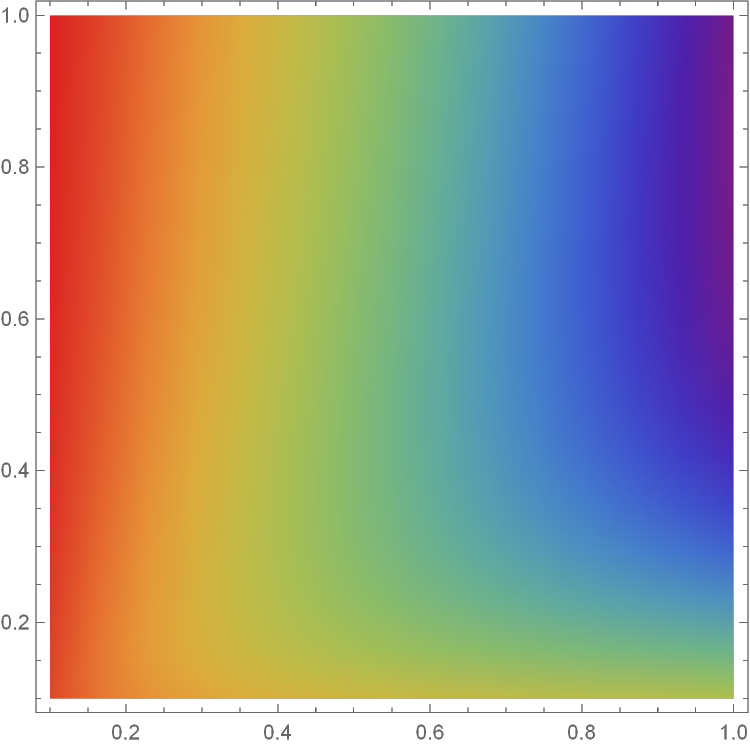} \hfill
\includegraphics[scale=0.30, valign=m]{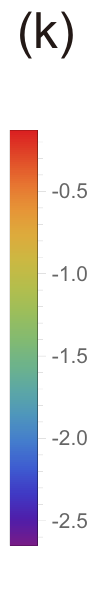} \\

%\noalign{\smallskip}
\vspace{0.2cm}
\rotatebox[origin=c]{90}{\scriptsize{$\tau_h = [0, 1]$; $\tau_l = [0, 10]$}}
\includegraphics[scale=0.25, valign=m]{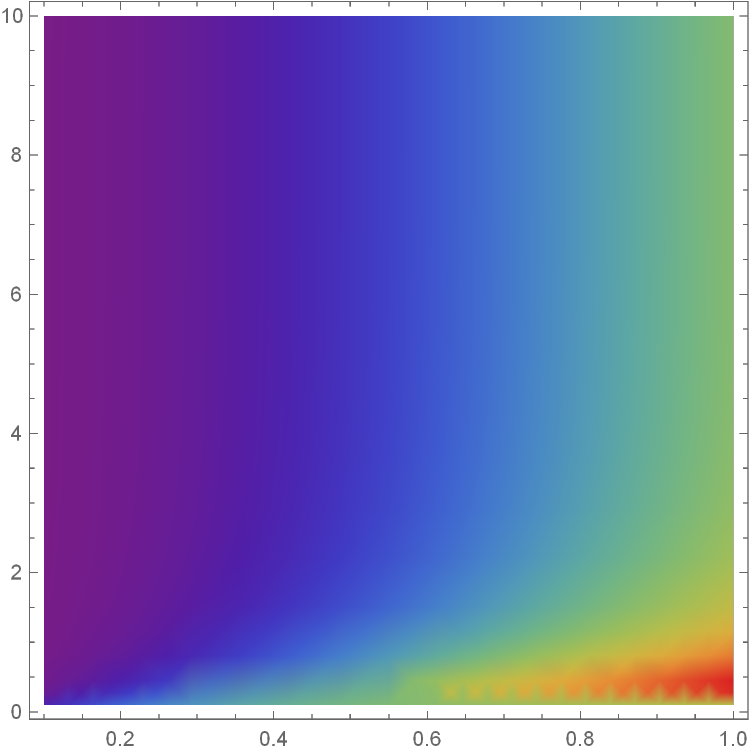} \hfill
\includegraphics[scale=0.30, valign=m]{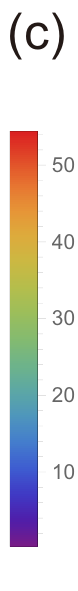} \hfill
&\includegraphics[scale=0.25, valign=m]{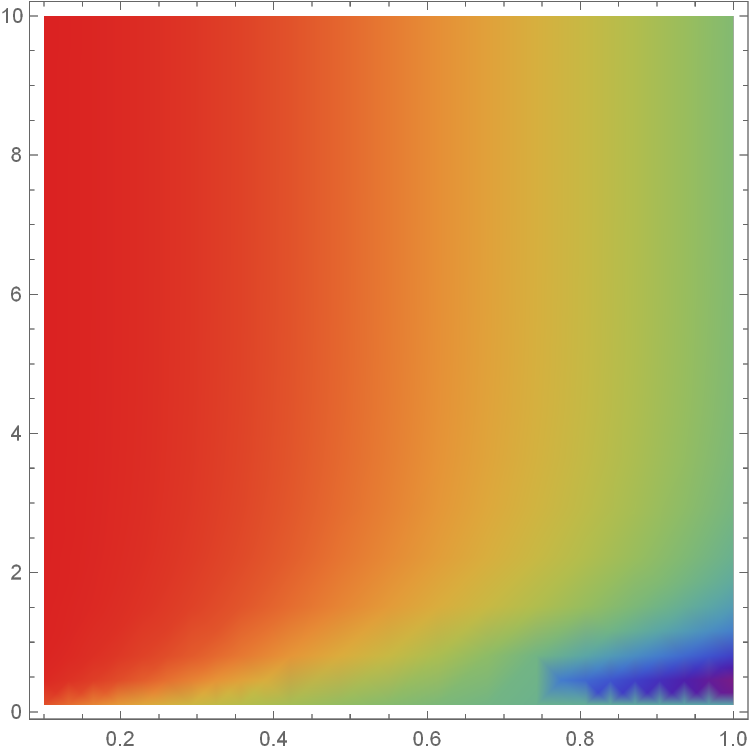} \hfill
\includegraphics[scale=0.30, valign=m]{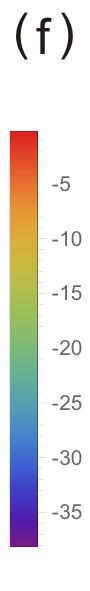} \hfill
&{}
&\includegraphics[scale=0.25, valign=m]{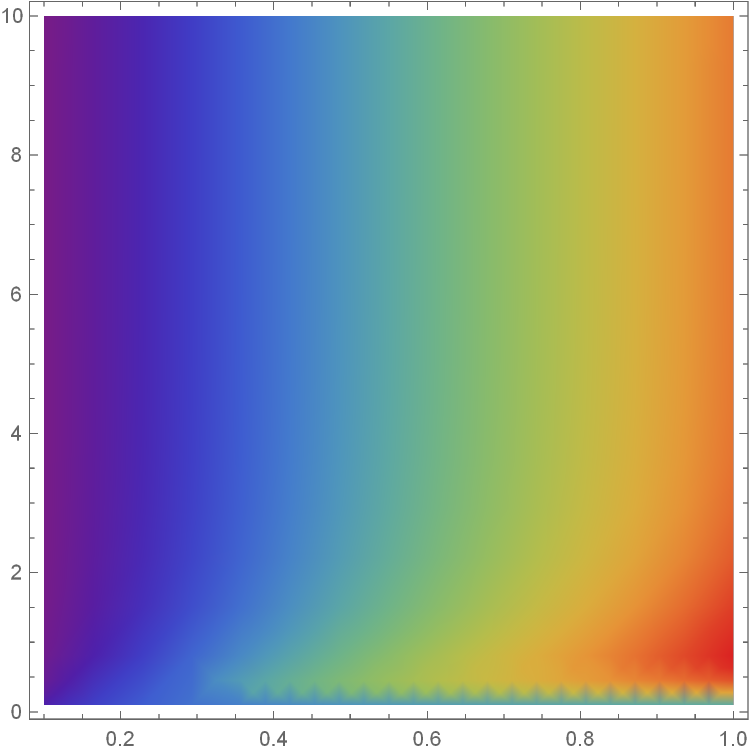} \hfill
\includegraphics[scale=0.30, valign=m]{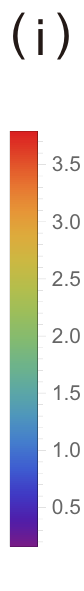} \hfill
&\includegraphics[scale=0.25, valign=m]{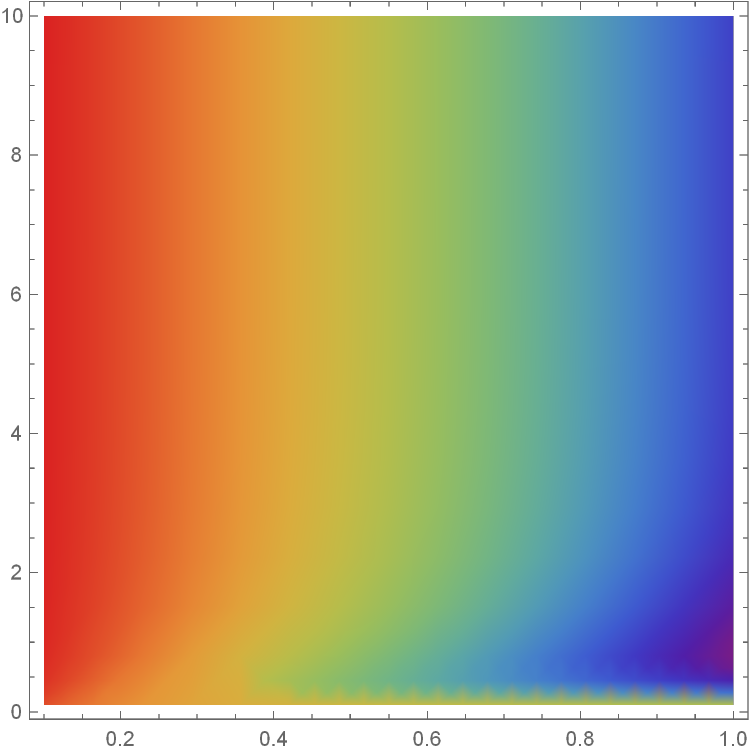} \hfill
\includegraphics[scale=0.30, valign=m]{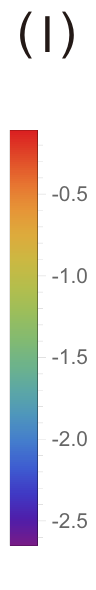} \\
\noalign{\smallskip}\\
\multicolumn{2}{c}{$n = 3$} &{} &\multicolumn{2}{c}{$n \rightarrow \infty$} \\ 
$\Delta S^{*}_{heating}$ &$\Delta S^{*}_{cooling}$ & &$\Delta S^{*}_{heating}$ &$\Delta S^{*}_{cooling}$ \\
\noalign{\smallskip}
\vspace{0.2cm}
\rotatebox[origin=c]{90}{\scriptsize{$\tau_h = [0, 10]$; $\tau_l = [0, 1]$}}
\includegraphics[scale=0.25, valign=m]{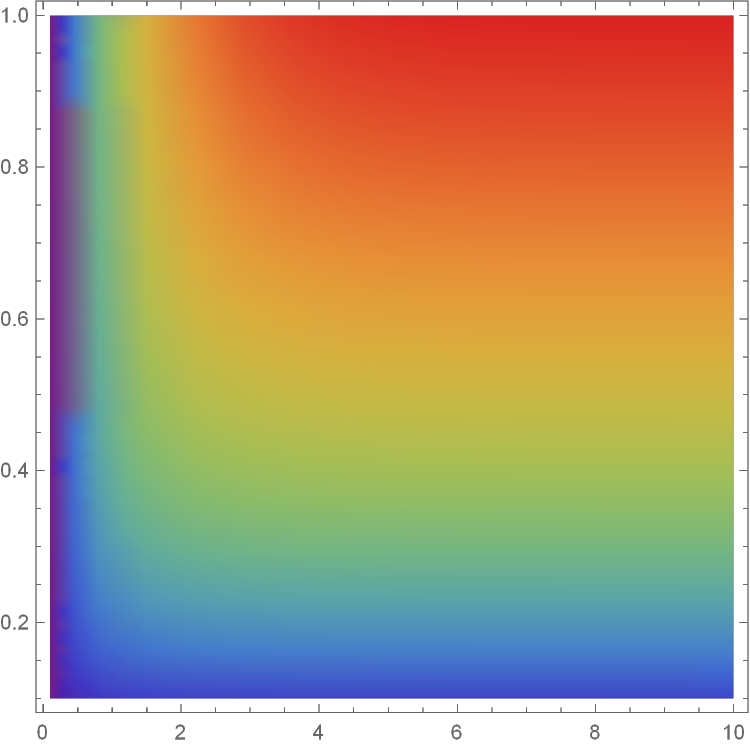} \hfill
\includegraphics[scale=0.30, valign=m]{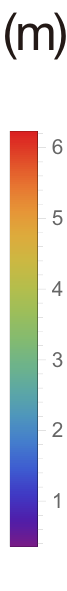} \hfill
&\includegraphics[scale=0.25, valign=m]{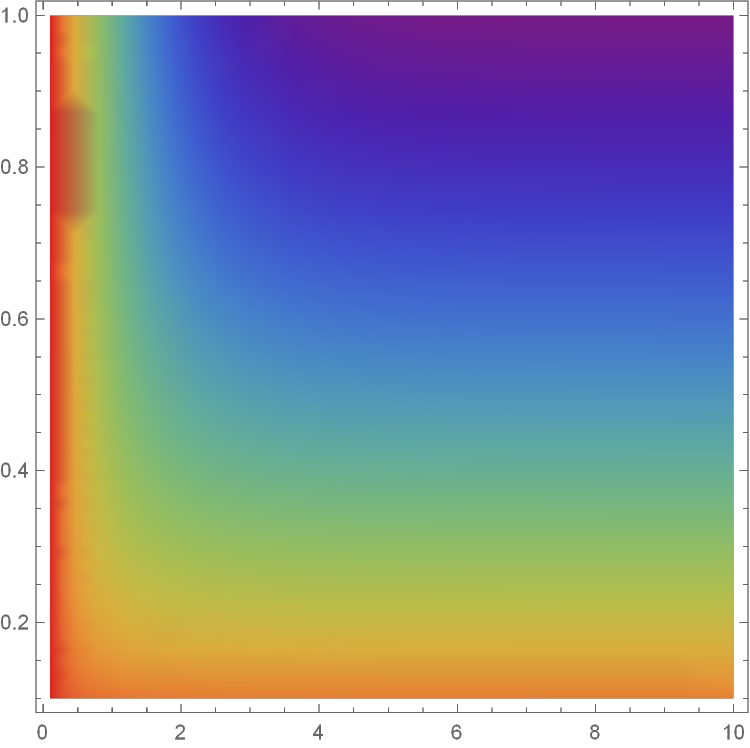} \hfill
\includegraphics[scale=0.30, valign=m]{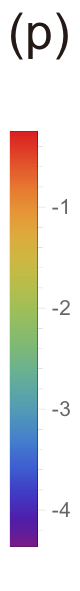} \hfill
&{}
&\includegraphics[scale=0.25, valign=m]{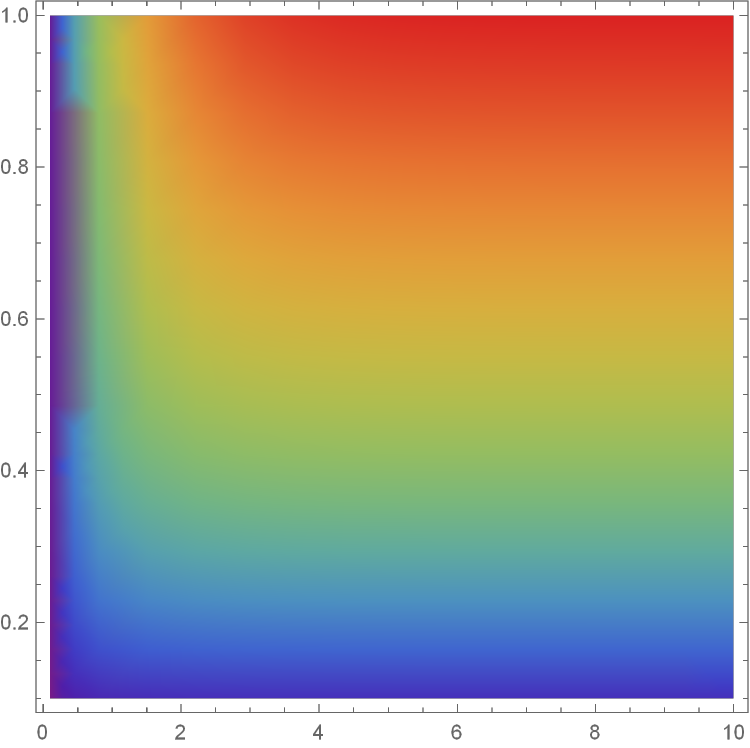} \hfill
\includegraphics[scale=0.30, valign=m]{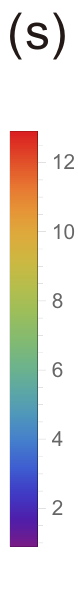} \hfill
&\includegraphics[scale=0.25, valign=m]{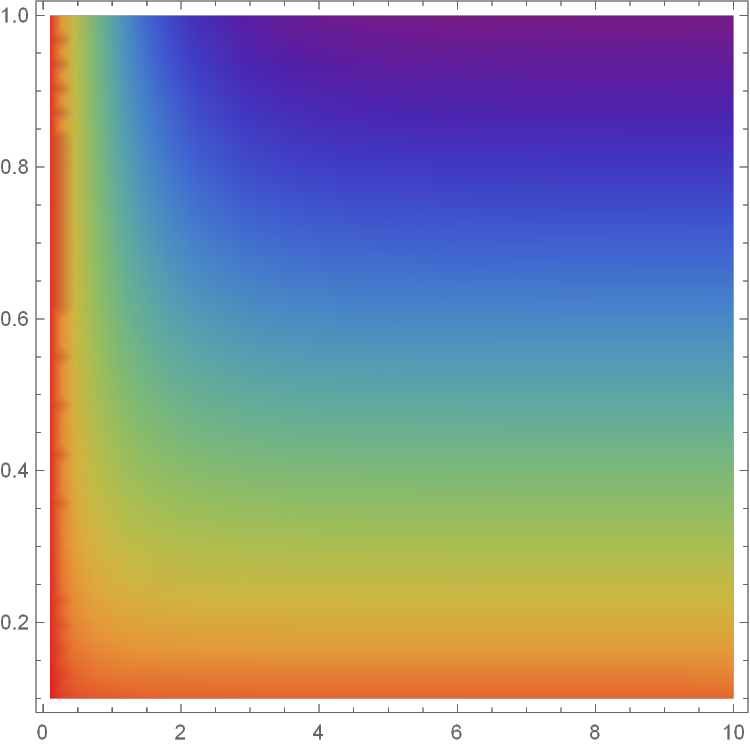} \hfill
\includegraphics[scale=0.30, valign=m]{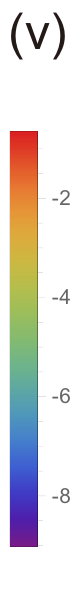} \\

%\noalign{\smallskip}
\vspace{0.2cm}
\rotatebox[origin=c]{90}{\scriptsize{$\tau_h = [0, 1]$; $\tau_l = [0, 1]$}}
\includegraphics[scale=0.25, valign=m]{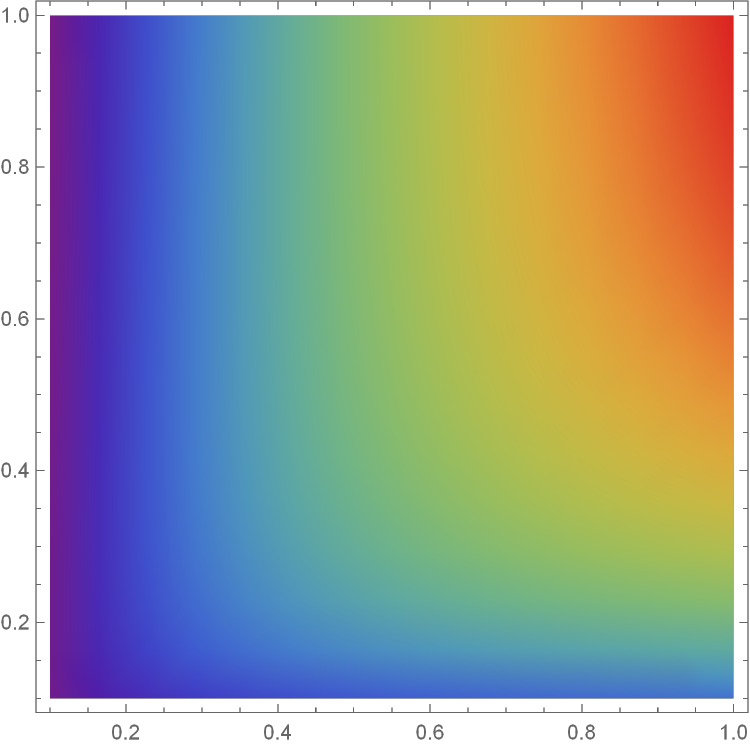} \hfill
\includegraphics[scale=0.30, valign=m]{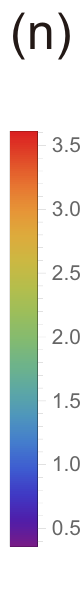} \hfill
&\includegraphics[scale=0.25, valign=m]{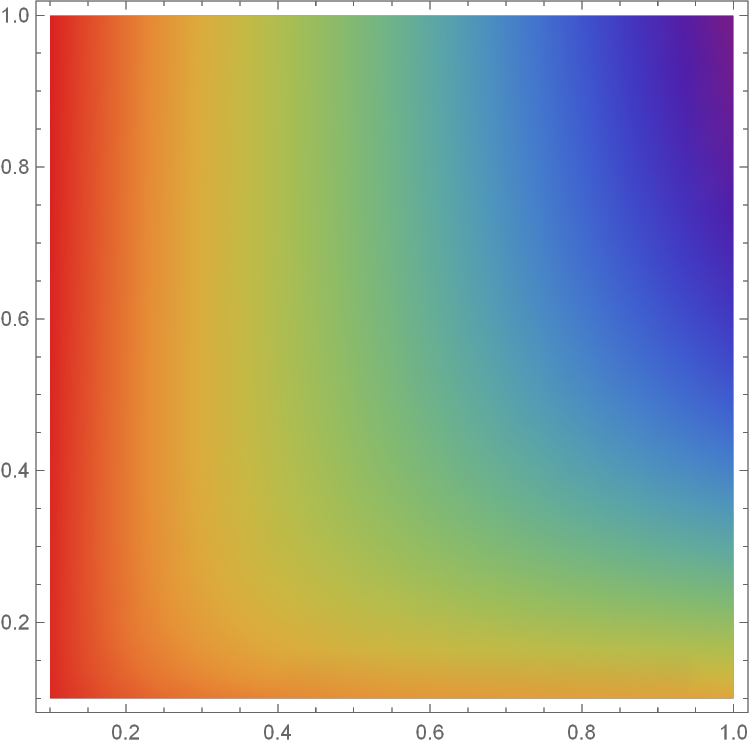} \hfill
\includegraphics[scale=0.30, valign=m]{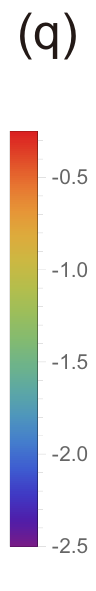} \hfill
&{}
&\includegraphics[scale=0.25, valign=m]{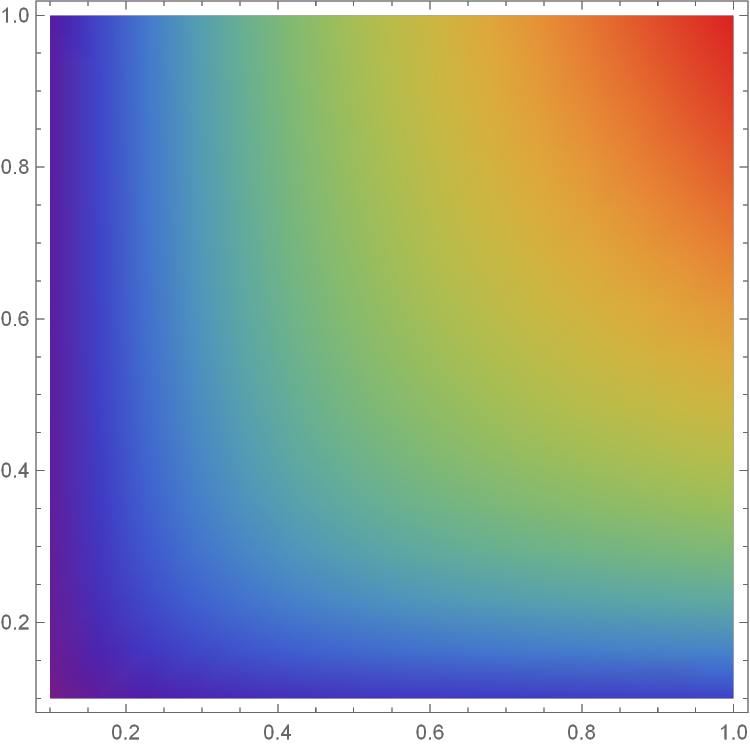} \hfill
\includegraphics[scale=0.30, valign=m]{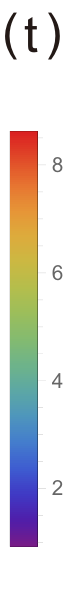} \hfill
&\includegraphics[scale=0.25, valign=m]{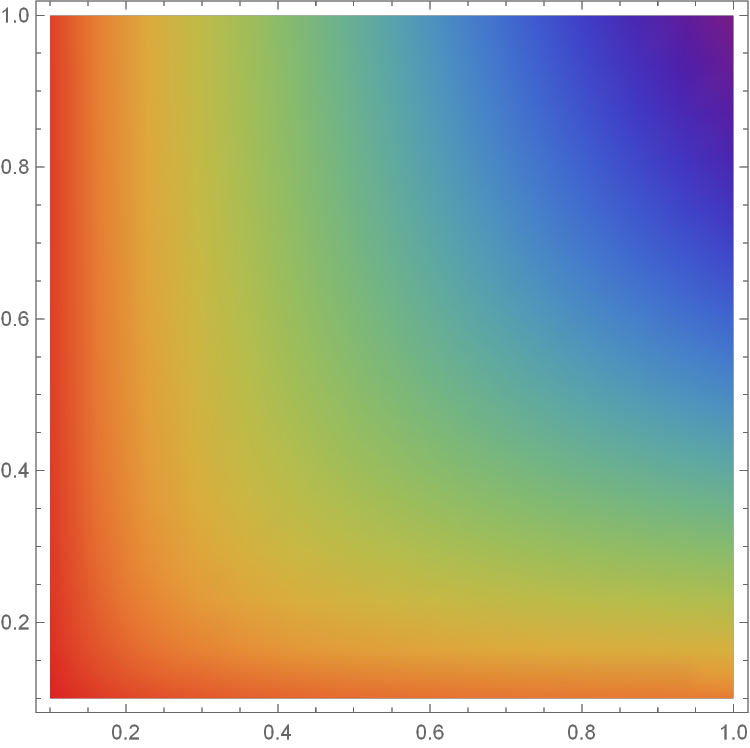} \hfill
\includegraphics[scale=0.30, valign=m]{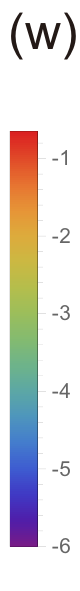} \\

%\noalign{\smallskip}
\vspace{0.2cm}
\rotatebox[origin=c]{90}{\scriptsize{$\tau_h = [0, 1]$; $\tau_l = [0, 10]$}}
\includegraphics[scale=0.25, valign=m]{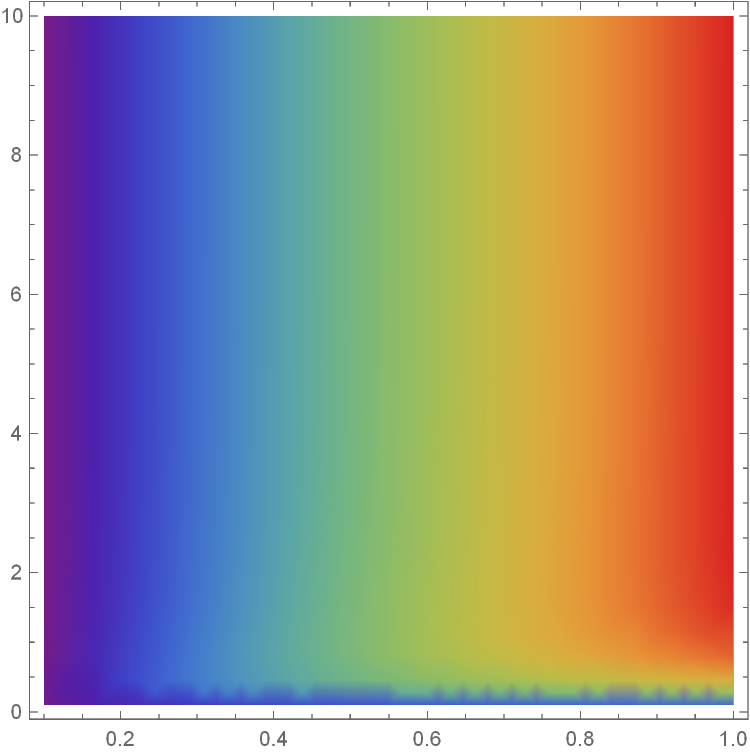} \hfill
\includegraphics[scale=0.30, valign=m]{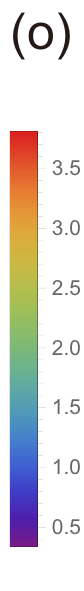} \hfill
&\includegraphics[scale=0.25, valign=m]{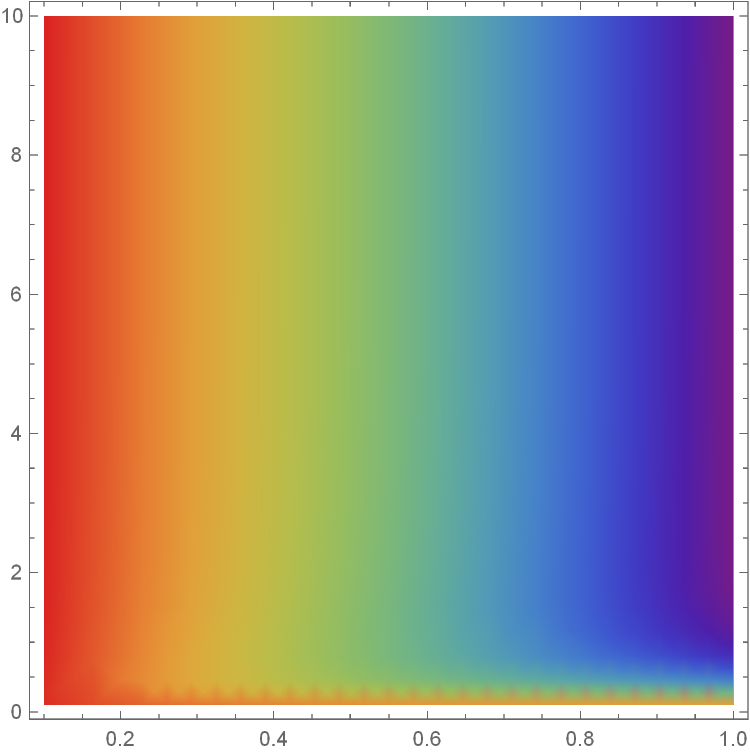} \hfill
\includegraphics[scale=0.30, valign=m]{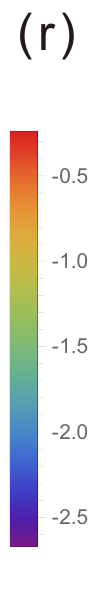} \hfill
&{}
&\includegraphics[scale=0.25, valign=m]{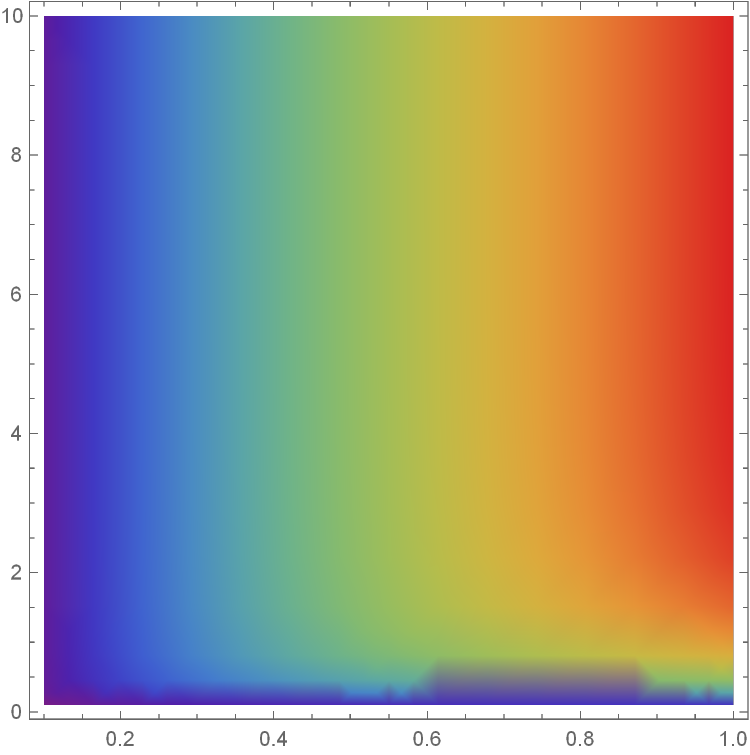} \hfill
\includegraphics[scale=0.30, valign=m]{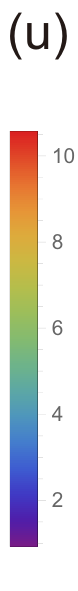} \hfill
&\includegraphics[scale=0.25, valign=m]{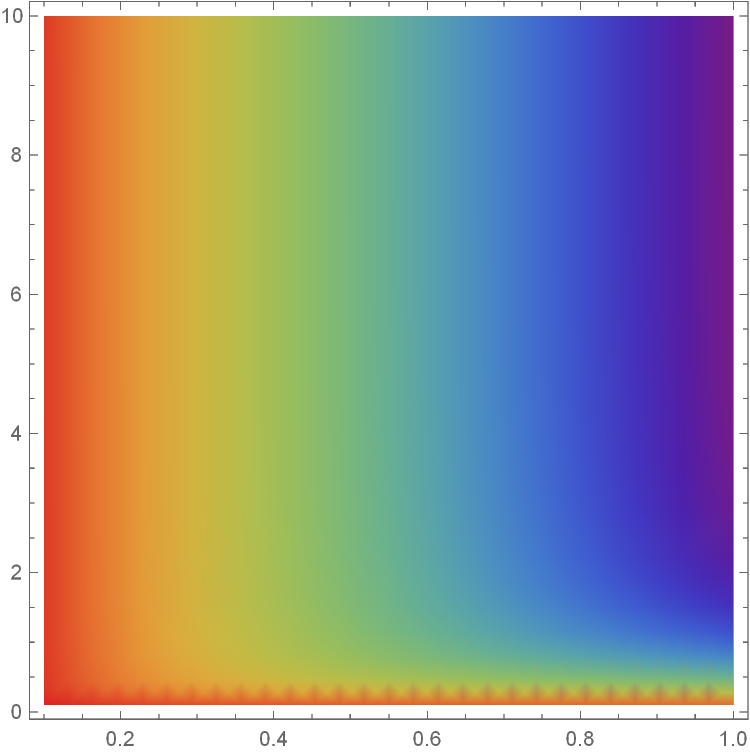} \hfill
\includegraphics[scale=0.30, valign=m]{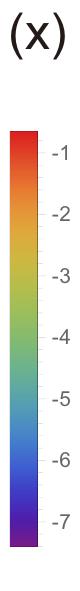} \\
\end{tabular}
\centering
\caption{\footnotesize\label{fig9}{The entropy change in heating (Eq. \ref{eq37}) and cooling (Eq. \ref{eq40}) process as a function of $\tau_h$ ($x$-axis) and $\tau_l$ ($y$-axis), with the temperatures $T_h$ and $T_l$ correspond to the interesting values in Figure \ref{fig5} with the same parameter for each $n$.}}
\end{figure*}

Entropy production is closely related to the irreversibility of the engine \cite{Mihaescu.2024}. The greater the entropy production, the more irreversible the engine becomes, leading to a decrease in efficiency at maximum power\cite{PhysRevA.99.062103}. Here, we visualize the entropy production by investigating the entropy change during the heating and cooling process. By deriving Eq. \ref{eq11} and utilizing the relations in Eqs. \ref{eq16}, \ref{eq21}, and \ref{eq25}, we obtained the entropy change in the heating process as follows
\begin{align}
   \Delta S_{heating}= \space{}
   &(b+1)A'a^{3}_{l}k_{B}^{b+1}\zeta(b+1) \times \nonumber \\ 
   &\Bigl[\left[T_{h}+\left(T_{2}-T_{h}\right)e^{-\alpha_{h}\tau_{h}}\right]^{b}-T_{2}^{b}\Bigr]
   \label{eq36}
\end{align}
which can be written in the form 
\begin{align}
   \Delta S^{*}_{heating}=\left[T_{h}+\left(T_{2}-T_{h}\right)e^{-\alpha_{h}\tau_{h}}\right]^{b}-T_{2}^{b} 
   \label{eq37}
\end{align}
with $\Delta S^{*}_{heating}=\Delta S_{heating}/C$, whereas
\begin{align}
   T_{2}=\frac{T_{l}e^{\alpha_{h}\tau_{h}} \left(e^{\alpha_{l}\tau_{l}}-1\right)+T_{h}\kappa^{\frac{1}{b}}\left(e^{\alpha_{h}\tau_{h}}-1\right)}{\kappa^{\frac{1}{b}}\left(e^{\alpha_{l}\tau_{l}+ \alpha_{h}\tau_{h}}-1\right)}
   \label{eq38}
\end{align}
Meanwhile, the entropy change within the cooling process is described as
\begin{align}
   \Delta S_{cooling}= \space{}
   &(b+1)A'a^{3}_{h}k_{B}^{b+1}\zeta(b+1) \times \nonumber \\ 
   &\Bigl[\left[T_{l}+\left(T_{4}-T_{l}\right)e^{-\alpha_{l}\tau_{l}}\right]^{b}-T_{4}^{b}\Bigr]
   \label{eq39}
\end{align}
with a similar idea, we rewrite the Eq. \ref{eq39} in the form
\begin{align}
   \Delta S^{*}_{cooling}=\left[T_{l}+\left(T_{4}-T_{l}\right)e^{-\alpha_{l}\tau_{l}}\right]^{b}-T_{4}^{b}  
   \label{eq40}
\end{align}
while
\begin{align}
   T_{4}=\frac{T_{l}\left(e^{\alpha_{l}\tau_{l}} -1\right)+ T_{h}\kappa^{\frac{1}{b}} \left(e^{\alpha_{h}\tau_{h}}-1\right) e^{\alpha_{l}\tau_{l}}}{e^{\alpha_{l}\tau_{l} +\alpha_{h}\tau_{h}}-1}
   \label{eq41}
\end{align}

In the cooling process, heat is transferred from the medium to the cold reservoir, so the change in entropy is negative. In finite time cycles such as endoreversible cycles, the entropy production during heating and cooling does not cancel each other or $\Delta S_{heating}+\Delta S_{cooling} > 0$, as shown in Figure \ref{fig9}. The amount of heat flow during heating stroke is not the same as the amount of heat flow during cooling stroke. Only reversible engines, such as Carnot engine, produce zero entropy during the cycle \cite{PhysRevA.99.062103}. Thus, the efficiency of an irreversible engine will never reach Carnot's efficiency. However, this entropy production could be minimized by reducing the amount of heat transferred to the medium or reducing the heating stroke time. According to Figure \ref{fig9}c, the entropy of heating process is relatively lower compared to Figures \ref{fig9}a and \ref{fig9}b; this result is noticed obviously by the presence of the color more "blue". These results are also in agreement with the results obtained in Figure \ref{fig8}b, in which EMP is relatively higher when the heating time $\tau_{h}$ is shortened.

Moreover, based on Figure \ref{fig8}b, both $\tau_{l}$ and $\tau_{h}$ are actually affected EMP. It can be seen that the blue and black dashes work at the same $\tau_{h}$, but EMP is higher at shorten $\tau_{h}$. Physically, the temperature of medium in blue dash is lower than temperature in black dash because more heat is injected from medium into the cold reservoir. In BEC, the fraction of condensed atoms depends on temperature (Eq. \ref{eq9}), the fraction of bosons condensed on the blue dash will be larger than on the black dash. When condensations are formed, these bosons will be occupied to the lowest quantum state or, macroscopically, the wave properties of each atom will collapse and interfere constructively with each other and form a single wave function \cite{ketterle1997coherence}. This coherence effect arises from the occupation of atoms in the condensate in the same quantum state,  which is described by the single wave function \cite{donley2002atom}. Based on this fact, blue will be "more coherent" than black. The increasing coherence in the boson state emerges when an incomplete thermalization process occurs, particularly when $\tau_h < \tau_l$, which leads to the reduction in entropy production (see Figure \ref{fig9}c). Thus, it can be said that the presence of this coherence due to incomplete thermalization contributes to an elevation in EMP \cite{PhysRevA.99.062103}. On the other hand, when $\tau_h > \tau_l$, this process does not guarantee an increase in coherence and a reduction in entropy. Therefore, we gain a specific case to enhance the EMP only when incomplete thermalization occurs at $\tau_h < \tau_l$ (see Figure \ref{fig8}b).

The study shows EMP is higher when using BEC as a working medium than normal boson gas (non-condensed conditions). The best performance is also given by the potential with $n=1$ because it produces the highest power and EMP. In addition, its critical temperature is relatively easy to reach and close to the temperature of experimental realizations of BEC itself \cite{bradley1995evidence}. Nevertheless, the power generated by $n=1$ is still very small when compared to the real engine due to the engine operating at very low temperatures even though the efficiency of using BEC is much higher than the efficiency provided by classical gas \cite{Erbay.1997,CA.1975,Deffner.2018,Leff.1987}. Since the power is determined by changes in internal energy during expansion and compression, it is possible to boost the power by adding more particles or by increasing external potential $\epsilon_{0}$ (in this calculation, we used $\epsilon_{0}$ corresponding to prior study \cite{Myers.2022}). However, increasing the number of particles will increase the density of the gas, so the interaction between atoms in gas cannot be neglected. Moreover, based on experimental results, BEC only occurs at very low densities \cite{Davis.1995,Anderson.2008,PhysRevLett.87.130402}. The critical temperature also depends on particle density, the rise in density also rising $T_{c}$ which is relatively easy to access \cite{Reppy.2000}. It is necessary to consider other physical aspects such as the volume of the potential as shown in Equation \ref{eq1} so that BEC can still occur.

\section{Conclusion} \label{sec5}
Bose-Einstein Condensation (BEC) is one of the matters which can be harnessed as a working medium in the QHE idea. Here, we focus on investigating EMP using BEC trapped in a generalized external potential instead of the non-BEC because of its dependent on medium properties and external potential. At the same range of $T$, for various degrees of potential $n$, EMP decreases as $n$ increases, whilst $\eta$ increases as $n$ increases. Moreover, we also investigate the EMP in BEC regimes by varying isochoric stroke times ($\tau_{l}$ and $\tau_{h}$). Both $\tau_{l}$ and $\tau_{h}$ directly affect the final temperatures of the medium in each isochoric stroke through the Fourier Conduction Law ($T_{1}$ and $T_{3}$) (see Equation \ref{eq21}). Despite the fact that EMP is decreasing for a longer stroke time, there is no significant difference for the same value of $\tau_{l}$ and $\tau_{h}$. However, if we set $\tau_{l}$ and $\tau_{h}$ properly, e.g., short stroke time in the heating isochoric process and long stroke time in the cooling isochoric process, the engine produces higher EMP than the one operates in long heating stroke time and short cooling stroke time, surprisingly also higher than operates in equal heating and cooling stroke time. This indicates that the setting of the stroke time of the engine can restrain the entropy production as well as quantum friction in the system. When the entropy change in the heating process exceeds the entropy in the cooling process, resulting in a positive $\Delta S$, residual coherence emerges due to incomplete thermalization. Furthermore, with the uniqueness of the BEC regime, the optimal work extracted can be obtained within the full excited expansion stroke, (the closest by $T_{C}$), and full condensed compression stroke, (the farthest by $T_{C}$). Hence, we conclude that the quantum Otto engine operates more effectively on potential $n=1$ than other variations of $n$ whose the highest critical temperature, so as being the highest EMP and highest power output producer. In-depth exploration for further study involves examining the interaction between the number of particles and the specific parameter $\epsilon_{0}$ with the external potential. Additionally, gaining a comprehensive insight into the interactions between bosonic particles significantly enhances our understanding, unveiling a more realistic behavior of these particles, especially under high-density conditions. These investigations hold paramount importance in the development of a genuinely realistic nano-engine, providing deeper insights into the dependencies governing its performance.

\begin{acknowledgments}
TEPS thanks the Faculty of Mathematics and Natural Sciences, Andalas University, for financially supporting this research with research grant No. 04/UN.16.03.D/PP/FMIPA/2022.
\end{acknowledgments}

%%\section*{Declarations}

%%The author declares that there is no conflict of interest regarding the publication of this manuscript. In addition, the ethical issues, including plagiarism, informed consent, misconduct, data fabrication and/or falsification, double publication and/or submission, and redundancies, have been completely observed by the authors.

\bibliography{main}% Produces the bibliography via BibTeX.

\end{document}